\documentclass[12pt]{article}

\usepackage[superscript,biblabel,nomove]{cite}

\usepackage{graphicx}
\usepackage{amsmath}
\usepackage{times}
\usepackage[bookmarks=false]{hyperref}
\hypersetup{
    colorlinks=true,
    linkcolor=blue,
    filecolor=magenta,      
    urlcolor=blue,
    citecolor=blue,
}
\usepackage{fancyhdr}

\newcommand{\eps}{\varepsilon}

\newenvironment{sciabstract}{%
\begin{quote} \bf}
{\end{quote}}

\renewcommand{\figurename}{Fig.}

\title{Network experiment demonstrates\\ converse symmetry breaking}

\author{
\hspace{-6mm}Ferenc~Molnar,$^{1}$
Takashi Nishikawa,$^{1,2}$
Adilson E. Motter$^{1,2}$\\
\\
\hspace{-6mm}\normalsize{$^{1}$Department of Physics and Astronomy, Northwestern University, Evanston, IL 60208, USA}\\
\hspace{-6mm}\normalsize{$^{2}$Northwestern Institute on Complex Systems, Northwestern University, Evanston, IL 60208, USA}%
}

\date{}

\begin{document} 

\maketitle 

\baselineskip18pt

\begin{sciabstract}
Symmetry breaking---the phenomenon in which the symmetry of a system is not inherited by its stable states---underlies pattern formation, superconductivity, and numerous other effects.  Recent theoretical work has established the possibility of \textit{converse} symmetry breaking (CSB), a phenomenon in which the stable states are symmetric only when the system itself is not. This includes scenarios in which interacting entities are required to be non-identical in order to exhibit identical behavior, such as in reaching consensus.  Here we present an experimental demonstration of this phenomenon.  Using a network of alternating-current electromechanical oscillators, we show that their ability to achieve identical frequency synchronization is enhanced when the oscillators are tuned to be suitably non-identical and that CSB persists for a range of noise levels.  These results have implications for the optimization and control of network dynamics in a broad class of systems whose function benefits from harnessing uniform behavior.
\end{sciabstract}



\vfill\begin{center}
Nature Physics \textbf{16}, 351--356 (2020)\\[1mm]
{\small The published version is available online at:\\[-1mm] \url{http://dx.doi.org/10.1038/s41567-019-0742-y}}
\end{center}

\baselineskip18pt
\clearpage

Synchronization~\cite{Pikovsky:01}---perhaps the most widely studied phenomenon in network dynamics~\cite{rev3,rev5,Pecora:15}---has
been observed in many 
contexts,
including both natural systems (e.g., circadian clock cells~\cite{Yamaguchi:03,nat1,Lu:16}, ecological populations~\cite{Ranta:95,Schwartz:02}, 
human menstrual cycles~\cite{McClintock:71}, and 
crowds of pedestrians~\cite{Strogatz:05}) 
and engineered systems (e.g., Boolean logic gates~\cite{eng1}, semiconductor lasers~\cite{Fischer:06,eng2,Argyris:16}, 
electrochemical and nanomechanical oscillators~\cite{Kiss:02,eng3,Fon:17}, 
and power 
generators~\cite{Hill:06,Mot:13,Dorfler:13}).
Such observations are significant because they show that
approximately homogeneous dynamics
can emerge in heterogeneous populations.
Yet, until recently, the prevailing view had been that 
homogeneity in the dynamics is facilitated by increased homogeneity in the population.
This view has now changed with the theoretical discovery~\cite{TakashiAIS} that, in numerous systems, 
heterogeneity can be required for stable identical synchronization---even when the entities are identically coupled to the population.

The underlying phenomenon,
which we term {\it converse symmetry breaking},
can be elegantly described using the notion of symmetry---a fundamental property that can characterize a system and has deep implications for its dynamics~\cite{Okuda:91,Golubitsky:1999,Golubitsky:1988,Nicosia:13,Pecora:14,Whalen:15,Sorrentino:16,Zhang:17,Cho:17,Barrett:17}.
In contrast to the well-known phenomenon of symmetry breaking, in which symmetry in the system 
implies
broken symmetry in the stable states, converse symmetry breaking represents a scenario in which symmetry in the stable states 
implies
broken symmetry in the system.
The interaction networks of many real systems are invariant under node permutations and hence possess symmetries~\cite{MacArthur:08}.
Symmetry breaking in networks include important examples of chimera states~\cite{ch1,ch2,Chimera3,ch3,ch4,Martens:13,ChimeraXP3}, in which a broken-symmetry state with coexisting groups of synchronized and non-synchronized nodes is observed even though the system is symmetric.
Converse symmetry breaking, on the other hand, has been predicted for oscillator networks in which the phenomenon can be mediated, for example, by amplitude dynamics~\cite{TakashiAIS}, couplings internal to the oscillators~\cite{Yuanzhao17}, and interaction delays~\cite{Yuanzhao2018}.
However, unlike symmetry breaking, evidence for converse symmetry breaking has thus far remained theoretical.

In this Article, we present the first experimental demonstration of converse symmetry breaking, in which we account for noise and other realistic 
features.
Our experimental system is designed to allow for frequency synchronization and consists of alternating current  
(AC) electromechanical oscillators, which are identically coupled in order to isolate the effect of 
oscillator
heterogeneity from that of coupling heterogeneity.
We show that, within the precision of experimental measurements, the optimal stability of frequency synchronization  can be enhanced 
by making the values of a tunable parameter of the 
oscillators---the 
damping coefficient---suitably different from each other.
Our results indicate that we can harness converse symmetry breaking in optimizing network dynamics. 
In potential applications to networks in which synchronization is desirable, this would translate to controlling oscillator heterogeneity to enable enhanced stability and performance.

\begin{figure*}[t]
\center
\includegraphics[width=0.97\textwidth]{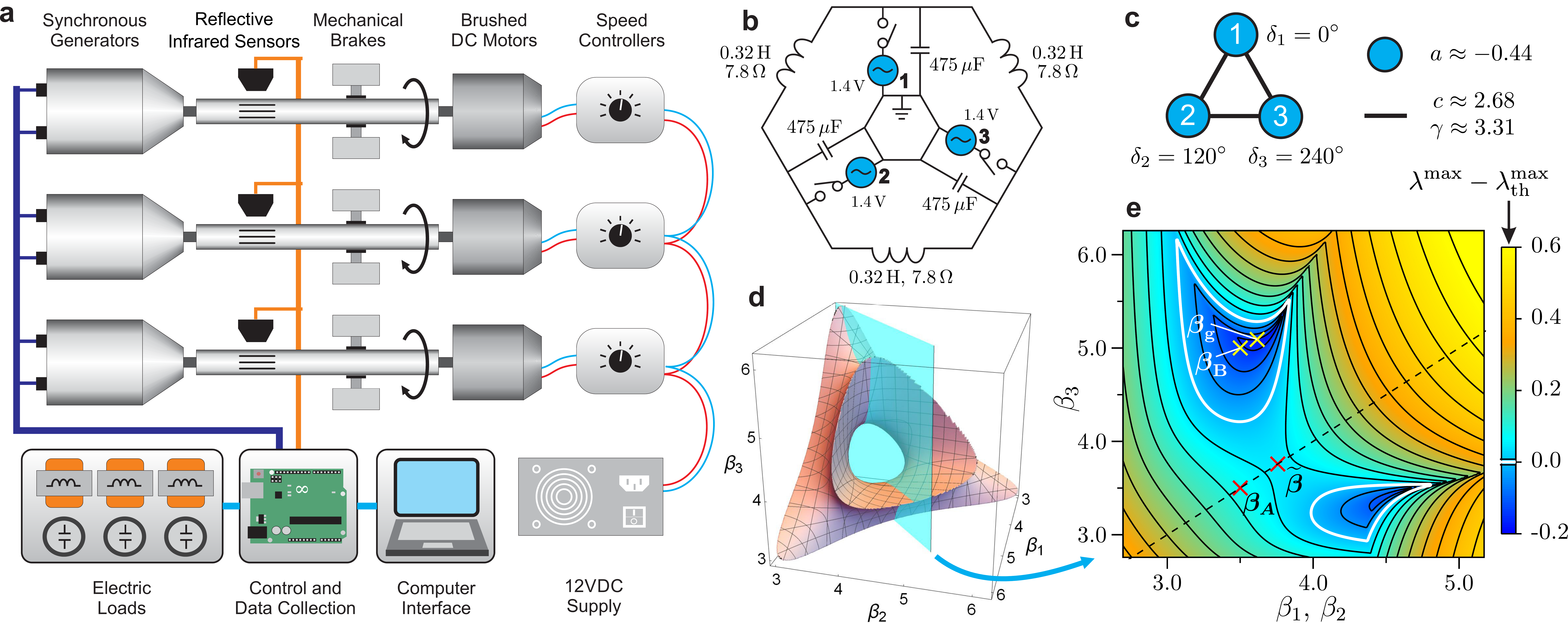}
\caption{\small 
{\bf Experiment involving a network of coupled electromechanical oscillators.}
{\bf a}, Main components of the 
experimental
setup,
including 
three
AC generators, 
three
DC motors driving them, and the computerized data acquisition system. 
{\bf b}, Diagram of the
AC electrical circuit connecting the three generators, running at $100$\,Hz.
{\bf c}, Network representation of the 
circuit, where the nodes represent generators and the links represent the
electrical interactions between them. 
The parameters 
characterizing 
the nodes and links are 
normalized by suitable references (i.e., given in per-unit quantities).
{\bf d}, Predicted stability  
of the frequency-synchronous splay states as a function of the generator parameters $\beta_i$. 
Inside the colored surface is 
the region of stability 
given by $\lambda^{\max} < \lambda^{\max}_\text{th}$ for noise corresponding to $\lambda^{\max}_\text{th} = -1.5$.
{\bf e}, Cross section of the stability landscape 
at the plane shown in {\bf d}.
Color-coded is the value of $\lambda^\mathrm{max}$ relative to $\lambda^{\max}_\text{th}$.
The optimal uniform assignment 
($\widetilde{\boldsymbol{\beta}}$) 
and the globally optimal non-uniform assignment ($\boldsymbol{\beta}_g$) are marked by red and yellow crosses, respectively.  Also marked by crosses are the projections of $\boldsymbol{\beta}_A$ and $\boldsymbol{\beta}_B$, the nearby assignments that we realize experimentally.}
\label{fig_setup}
\end{figure*}

Figure~\ref{fig_setup}a illustrates the main components of our experiment, in which three permanent-magnet 
generators are
mechanically driven by DC motors with adjustable speed and a separate 12V DC power supply.
The generators are chosen to have identical parameters (e.g., internal damping coefficient, internal impedance, and terminal voltage at various speeds) within manufacturing precision; see 
Methods for details.
To allow for heterogeneous configurations of the generators, their shafts are equipped with mechanical brakes that can be used to adjust friction.
The generators' output is connected to a set of electric loads (inductors and capacitors) 
forming the circuit depicted in Fig.~\ref{fig_setup}b.
The parameters of the circuit components are chosen for the system to be symmetric with respect to rotational permutations of the generators ($1 \to 2 \to 3 \to 1$).
The pattern of coupling among the generators can thus be represented as a rotationally symmetric network of three nodes (generators) connected by three identical links, each corresponding to a circuit equivalent to  
the so-called $\pi$ model~\cite{Grainger:1994}.
Each generator is fixed on its own platform to minimize mechanical coupling with the other generators via vibrations.

We focus on frequency-synchronous states of the system in which the voltage frequencies of the generators are all equal to a constant frequency $\omega_\mathrm{s}$.
On short timescales, of the order of $1$\,s or less, the 
dynamics of the generators when the system is close to such a state can be described by
a coupled oscillator model~\cite{And:03,Nish:15}.  When written for an arbitrary number $n$ of 
generators,
the model equation reads:
\begin{equation}\label{eq-en}
\ddot{\delta_i} + \beta_i \dot{\delta_i} = a_i - \sum_{k \neq i} c_{ik} \sin \left(\delta_i - \delta_k - \gamma_{ik} \right)
+ \eps\xi_i(t),
\end{equation}
where $\delta_i$ is the internal electrical angle for generator $i$ (a state variable related to the rotor shaft angle by a factor determined by the number of poles in the generator),
relative to a reference frame rotating at the synchronous frequency $\omega_\mathrm{s}$; 
the constant $\beta_i$ is an effective damping parameter (capturing both mechanical and electrical damping, normalized by the generator's inertia);
$a_i$ is a parameter representing the net power accelerating the generator's rotor (i.e., the mechanical power provided by the DC motor driving the generator, minus the power consumed by the network components and the power lost to damping);
$c_{ik}$ and $\gamma_{ik}$ are the coupling strength and phase shift characterizing the electrical interactions between the generators;
$\xi_i(t)$ is a random function representing dynamical noise; and $\eps$ is the noise amplitude
(see Supplementary Information, Sec.~\ref{app:eq-en-derivation} for a derivation of the deterministic part, Methods for details on modeling dynamical noise, and Supplementary Fig.~\ref{fig:validation} for validation).
In the co-rotating frame, the
frequency-synchronous states correspond to the fixed-point solutions 
of Eq.~\eqref{eq-en} with $\eps=0$, characterized by $\omega_i \equiv \dot{\delta}_i = 0$, $\forall i$.
Note that uniform angle shifts of one such solution represent the same state, as all angle differences remain unchanged.
The deterministic part of Eq.~\eqref{eq-en} has the same form as that used to model the dynamics of generators in power grids~\cite{Mot:13,Dorfler:13}, which have recently been studied extensively in the network dynamics 
community~\cite{pg3,pg2,Men:14,Auer:16,Schafer:18}, but here the equation is used to model coupled electromechanical oscillators that are tunable and not constrained to operating states of power grids.

Linearizing 
Eq.~\eqref{eq-en} for $\eps=0$ around a fixed point with $\delta_i = \delta_i^*$, we obtain 
$\dot{\mathbf{x}} = \mathbf{J}\mathbf{x}$, 
where $\mathbf{x} = \bigl( \begin{smallmatrix} \Delta\boldsymbol{\delta} \\ \Delta\boldsymbol{\omega} \end{smallmatrix}\bigr)$ and $\mathbf{J} = \bigl( \begin{smallmatrix}\,\,\,\,\mathbf{0} & \,\,\,\,\mathbf{1}\\ -\mathbf{P} & -\mathbf{B}\end{smallmatrix}\bigr)$.
Here, $\Delta\boldsymbol{\delta}$ and $\Delta\boldsymbol{\omega}$ are the $n$-dimensional vectors of angle and frequency
deviations, 
$\delta_i - \delta_i^*$ and $\omega_i - 0 = \omega_i = \dot{\delta}_i$, respectively.
The $n \times n$ matrix $\mathbf{P}=(P_{ik})$ is given by
\begin{equation}\label{eqn:P}
P_{ik} = \begin{cases}
- c_{ik} \cos (\delta_i^* - \delta_k^* - \gamma_{ik}), & i \neq k, \\
-\sum_{k' \neq i} P_{ik'}, & i = k,
\end{cases}\vspace{3mm}
\end{equation}
$\mathbf{B}$ is the $n \times n$ diagonal matrix with 
$\beta_i$ 
as its diagonal elements, and $\mathbf{0}$ and $\mathbf{1}$ denote the null and identity matrices of size $n$, respectively.
The stability of the corresponding frequency-synchronous state of Eq.~\eqref{eq-en} with $\eps=0$ is thus determined by the eigenvalues $\lambda_i$ of the Jacobian matrix $\mathbf{J}$, excluding the identically zero eigenvalue (which we denote by $\lambda_1$) present only because of the zero row-sum property of $\mathbf{P}$.
Specifically, if the maximal real part of these eigenvalues is negative, i.e., the Lyapunov exponent $\lambda^\mathrm{max} \equiv \max_{i \ge 2} \mathrm{Re} (\lambda_i)$ is negative, then the state is asymptotically stable,
and smaller $\lambda^\mathrm{max}$ implies stronger stability.
(The zero eigenvalue $\lambda_1$ is excluded because it is associated with perturbations that uniformly shift phases, which lead to equivalent fixed points corresponding to the same frequency-synchronous state.) The problem of maximizing this stability with respect to system parameters can thus be formulated as the optimization of $\lambda^\mathrm{max}$. This is similar to the problem of optimizing the largest real part of the eigenvalues of a matrix, known as the spectral abscissa (relevant when the matrix has no identically null eigenvalues) \cite{r1,r2,r3}, and the problem of optimizing the second largest eigenvalue of a Laplacian matrix, known as the algebraic connectivity \cite{r4,r5}. Some previous studies have considered problems concerning the optimization of damping parameters \cite{r2,r3}, but they do not focus on relations to system symmetry and typically exclude the class of non-positive definite non-symmetric coupling matrices relevant to the experiment considered here. 

In the presence of dynamical noise (i.e., when $\eps>0$), we can show that, for general classes of discrete- and continuous-time noise models, there is a (negative) threshold $\lambda^\mathrm{max}$ value for the stability of the frequency-synchronous state.
We denote this stability threshold by $\lambda^{\max}_\text{th}$; the frequency synchronization is stable below this threshold and unstable above it (see Methods for details).

\begin{figure*}[t!]
\center
\includegraphics{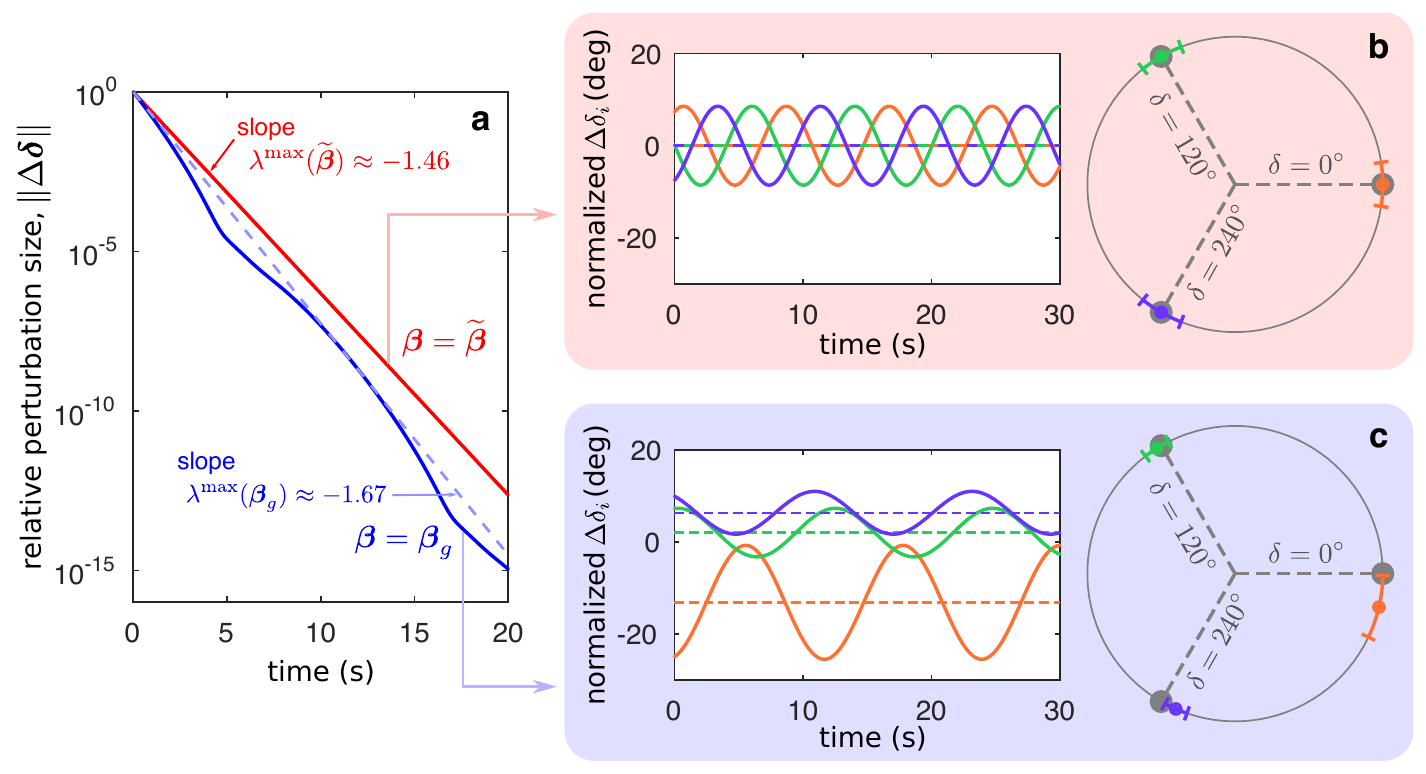}
\caption{\small
{\bf Oscillator heterogeneity breaks the symmetry of the dominant eigenmodes.}
The predicted slowest decaying eigenmodes of deviations from the splay state are visualized for the system corresponding to  Fig.~\ref{fig_setup}b--e.
{\bf a}, Magnitude $\|\Delta\boldsymbol{\delta}\|$ of the eigenmodes, which decay at an overall exponential rate $\lambda^\mathrm{max}(\boldsymbol{\beta})$, for $\boldsymbol{\beta} = \widetilde{\boldsymbol{\beta}}$ (red) and for $\boldsymbol{\beta} = \boldsymbol{\beta}_g$ (blue).
For $\boldsymbol{\beta} = \widetilde{\boldsymbol{\beta}}$, the decay shown is for a combination of two oscillatory
eigenmodes associated with complex conjugate eigenvalues corresponding to $\lambda^\mathrm{max}(\widetilde{\boldsymbol{\beta}})$, whereas for $\boldsymbol{\beta} = \boldsymbol{\beta}_g$, the decay is for a combination of three eigenmodes
corresponding to $\lambda^\mathrm{max}(\boldsymbol{\beta}_g)$, of which two
oscillatory modes are associated with complex conjugate eigenvalues and one non-oscillatory
mode is associated with a real eigenvalue.
{\bf b}, Dynamics of individual oscillators (orange, green, and purple for oscillator $i=1,2,3$, respectively) given by the eigenmodes for $\boldsymbol{\beta} = \widetilde{\boldsymbol{\beta}}$, after normalization that removes the exponential decay shown in {\bf a}.  (Left) Dynamics with respect to time.  (Right) Amplitude of eigen-oscillations indicated by the bars drawn along the unit circle (normalized such that $\|\Delta\boldsymbol{\delta}\|=0.5$). The gray dots in the background represent the splay state, in which $\delta_i$ for different $i$ are 120$^\circ$ apart from each other.
{\bf c}, Same as in {\bf b}, but for $\boldsymbol{\beta} = \boldsymbol{\beta}_g$.  In this case, the bars for perturbation amplitudes are shifted by offsets corresponding to the non-oscillatory eigenmode.
We observe that the dominant eigenmodes are rotationally symmetric for $\boldsymbol{\beta} = \widetilde{\boldsymbol{\beta}}$, but this symmetry is broken for $\boldsymbol{\beta} = \boldsymbol{\beta}_g$.
For an animated version of the plots in {\bf b} and {\bf c}, see \url{https://youtu.be/R_BOIWXYtSk}
}
\label{jacobian_asym_figure}
\end{figure*}

We now use our stability analysis to 
derive a condition for observing converse symmetry breaking in this system.
Note that $\lambda^\mathrm{max}$ is a function of $\boldsymbol{\beta} = (\beta_1,\ldots,\beta_n)$, and that, 
even though Eq.~\eqref{eq-en} does not depend on $\beta_i$ when restricted to the synchronous states $\dot{\delta}_i = 0$, the corresponding variational equation $\dot{\mathbf{x}} = \mathbf{J}\mathbf{x}$ does.
This implies that making $\beta_i$ heterogeneous, which breaks the symmetry of the system, generally leads to symmetry breaking in the eigenmodes around the synchronous state (Fig.~\ref{jacobian_asym_figure}). 
We see that, if two values of $\boldsymbol{\beta}$ correspond to distinct values of $\lambda^\mathrm{max}$, then there is a range of noise intensities (and thus of $\lambda^{\max}_\text{th}$) for which the state is stable for one $\boldsymbol{\beta}$ value and unstable for the other (see 
Methods
for details).
Thus, a condition for exhibiting converse symmetry breaking is that there exists $\boldsymbol{\beta}^*$ representing a non-uniform $\beta_i$ assignment for which
\begin{equation}\label{eqn:csb}
\lambda^\mathrm{max}(\boldsymbol{\beta}^*) < \min \{ 0, \lambda^\mathrm{max}(\widetilde{\boldsymbol{\beta}})\},
\end{equation}
where we define $\widetilde{\boldsymbol{\beta}} \equiv (\widetilde{\beta},\ldots,\widetilde{\beta})$ to represent the (uniform) $\beta_i$ assignment that minimizes $\lambda^\mathrm{max}$ under the constraint that $\beta_1 = \cdots = \beta_n$.

In the design of our experiment, the electrical parameters of the AC circuit (indicated in Fig.~\ref{fig_setup}b)
were chosen to ensure that 
the circuit is rotationally symmetric and the frequency-synchronous states that inherit that symmetry have 
$\lambda^\mathrm{max} < 0$
at frequency $\omega_\mathrm{s} = 100$\,Hz (see 
Methods
for details). 
The specific states we focus on are known as \textit{splay states}~\cite{splay}, which for a ring of $n$ phase oscillators are defined as states in which phase differences between consecutive oscillators are all equal to $2\pi/n$.
With the choice of parameters used in the experiment, Eq.~\eqref{eq-en} describes a reduced $3$-node network system with $a_i = a$, $c_{ik} = c$, and $\gamma_{ik} = \gamma$ (values given in Fig.~\ref{fig_setup}c), which has $3$-fold rotational symmetry (as the original AC circuit).
This system has two splay states (with identical $\lambda^\mathrm{max}$), corresponding to two fixed points of Eq.~\eqref{eq-en} for which the phase angles of the adjacent generators are exactly $120$ degrees apart (i.e., 
$\delta_2 - \delta_1 = \pm 120^\circ$ and $\delta_3 - \delta_1 = \mp 120^\circ$).
Each of these is a rotationally symmetric state of the system, since applying any rotational permutation of the generators would result in the same state (equivalent up to a uniform shift of all $\delta_i$).
In these states, the effective coupling pattern expressed in the matrix $\mathbf{P}$ takes 
a form that favors converse symmetry breaking (see 
Methods
for details).

The stability landscape---the function $\lambda^\mathrm{max}(\boldsymbol{\beta}) = \lambda^\mathrm{max}(\beta_1, \beta_2, \beta_3)$---for
the frequency-syn\-chronous splay states of our experimental system
exhibits the same rotational symmetry as the system itself, as illustrated in Fig.~\ref{fig_setup}d assuming ideal system components (see 
Methods
for details).
Due to 
this
symmetry, the landscape
has global minima (all with the same $\lambda^\mathrm{max}$ value)
at three different locations in the $\boldsymbol{\beta}$-space, related by a $120$-degree rotation of the $\beta_i$-axes.
One of these minima, which we denote by $\boldsymbol{\beta}_g$, is located approximately at 
$(3.69, 3.68, 5.18)$
and corresponds to a globally optimal (non-uniform) assignment of the generator parameters with 
$\lambda^\mathrm{max}(\boldsymbol{\beta}_g) \approx -1.67$. 
In contrast, the optimal uniform assignment (i.e., the optimum under the constraint $\beta_1 = \beta_2 = \beta_3$) corresponds to the point $\widetilde{\boldsymbol{\beta}} = (\widetilde{\beta},\widetilde{\beta},\widetilde{\beta})$ with $\widetilde{\beta} \approx 3.76$ and
$\lambda^\mathrm{max}(\widetilde{\boldsymbol{\beta}}) \approx -1.46$. 
These two points are marked on the cross section of the stability landscape shown in Fig.~\ref{fig_setup}e.
Since condition~\eqref{eqn:csb} is satisfied for $\boldsymbol{\beta}^* = \boldsymbol{\beta}_g$, the system 
is predicted to exhibit converse symmetry breaking: the rotationally symmetric, frequency-synchronous splay states are necessarily unstable for any uniform $\beta_i$ assignment (corresponding to a rotationally symmetric system), whereas they become stable for the non-uniform $\beta_i$ assignment in $\boldsymbol{\beta}_g$ (corresponding to a non-rotationally symmetric system) for a range of noise intensities.

Specific uniform and non-uniform $\beta_i$ assignments were implemented in our experiment by adjusting the frictional brakes on the generator shafts 
(see 
Supplementary Information, Sec.~\ref{app:eq-en-derivation}
on how friction relates to $\beta_i$).
Due to the physical limitations and finite measurement accuracy, the precision of a $\beta_i$ value that we could realize
was $\pm 0.1$ (see 
Supplementary Information, Sec.~\ref{supp-beta-meas}).
In addition, changes in the parameters of generator--motor units over time due to external effects (e.g., heating) as well as inherent heterogeneity of the system components (despite being manufactured to be identical) can distort the landscape itself and thus shift the locations of $\boldsymbol{\beta}_g$ and $\widetilde{\boldsymbol{\beta}}$.
Even though care was taken to minimize deviations from the designed values (see
Methods), it would be experimentally challenging to realize these points exactly.
Therefore, we considered two points on the landscape that we were able to realize and confirm with measurements: 
$\boldsymbol{\beta}_A \equiv (3.4\,\text{\scriptsize$\pm\,0.1$}, 3.6\,\text{\scriptsize$\pm\,0.1$}, 3.5\,\text{\scriptsize$\pm\,0.1$}) \approx \widetilde{\boldsymbol{\beta}}$
and 
$\boldsymbol{\beta}_B \equiv (3.4\,\text{\scriptsize$\pm\,0.1$}, 3.6\,\text{\scriptsize$\pm\,0.1$}, 5.0\,\text{\scriptsize$\pm\,0.1$}) \approx \boldsymbol{\beta}_g$
(projections of both points are marked in Fig.~\ref{fig_setup}e).
Our theoretical predictions for the stability at these points are $\lambda^\mathrm{max}(\boldsymbol{\beta}_A) \approx -1.42$ and $\lambda^\mathrm{max}(\boldsymbol{\beta}_B) \approx -1.63$, which confirms the property $\lambda^\mathrm{max}(\boldsymbol{\beta}_B) < \lambda^\mathrm{max}(\boldsymbol{\beta}_A)$ that is predicted to enable converse symmetry breaking in the experiment.

To provide experimental evidence for converse symmetry breaking, we
performed multiple experimental runs
while measuring the terminal voltage of each generator with a 
voltage sensor circuit
and concurrently processing the measurements (see
Methods for details). These measurements yielded multiple time series of terminal voltage phasors (angles and magnitudes) and the corresponding electrical angles $\delta_i$, both recorded at $3{,}320$ samples per second.
The total recorded time span was $\approx\!\!1.5$\,h each for the $\boldsymbol{\beta}_A$ and $\boldsymbol{\beta}_B$ configurations. 
For our analysis, we focused on
segments
of the time series in which the splay states were observed (allowing for a maximum of $\pm 10$ degree deviation in $\delta_i$; see 
Methods
for the full details on this criterion).
Since system parameters may fluctuate and shift gradually during an experimental run, we estimated a (slightly different) steady state for each segment using the averages of the measured phasor angles $\delta_i^*$ over all data points in the segment.
Using these steady-state values, as well as measured system parameters, we estimated $\lambda^\mathrm{max}$ for each segment of each $\boldsymbol{\beta}$ configuration (see 
Methods
for the procedure to calculate $\lambda^\mathrm{max}$).

\begin{figure*}[t]
\center
\includegraphics[width=\textwidth]{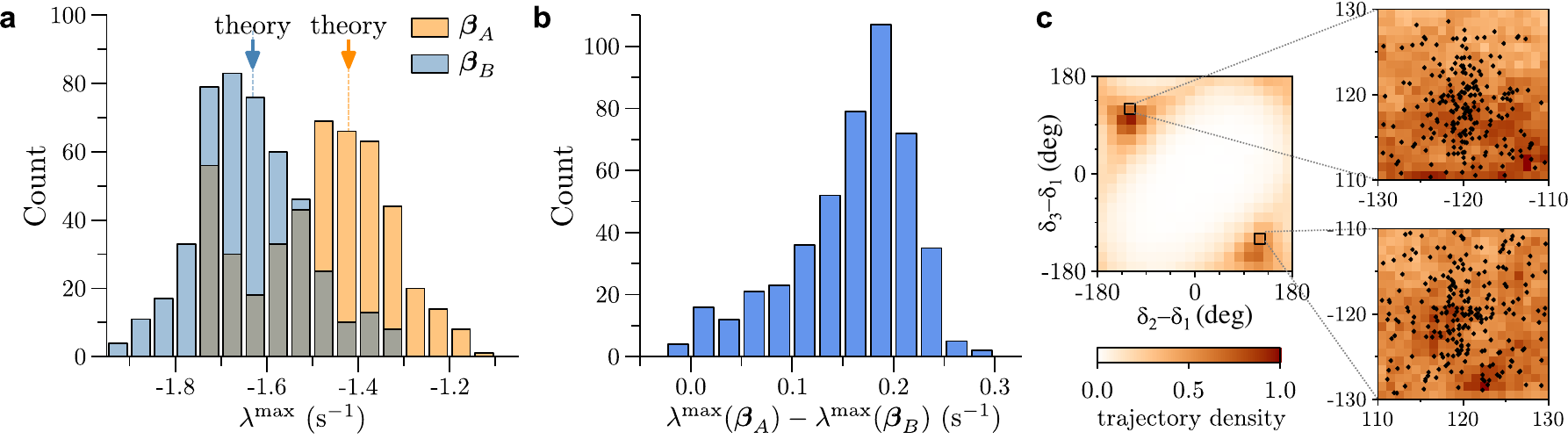}
\vspace{-5mm}
\caption{\small
{\bf Experimental confirmation of 
converse symmetry breaking.}
{\bf a}, Distribution of the Lyapunov exponents $\lambda^{\mathrm{max}}$ obtained from individual time-series segments for the uniform (${\boldsymbol\beta}_A$, orange histogram) and non-uniform (${\boldsymbol\beta}_B$, blue histogram) configurations.
For each segment, we used the measured splay state and associated parameters, and calculated $\lambda^{\mathrm{max}}$ as an average over that segment.
The downward arrows indicate the predicted values of $\lambda^{\mathrm{max}}$ for the theoretically calculated splay states (in which the angles are exactly 120 degrees apart).
{\bf b},~Distribution of synchronization stability improvement achieved by changing the generator parameters from ${\boldsymbol\beta}_A$ to ${\boldsymbol\beta}_B$ for the same states, as measured by the difference $\lambda^{\mathrm{max}}({\boldsymbol\beta}_A) - \lambda^{\mathrm{max}}({\boldsymbol\beta}_B)$. 
{\bf c}, Inferred splay states and density plot of time-series trajectories.
The horizontal and vertical axes are the phase angles $\delta_2$ and $\delta_3$ of the second and third generators, respectively, relative to the first generator. 
The trajectory density was estimated for each pixel using all measured time series.
The color scale is normalized to the highest measured density. 
The dots in the enlarged sections mark the steady splay states determined for the time-series segments we considered 
(275 and 190 
segments for the ${\boldsymbol\beta}_A$ and ${\boldsymbol\beta}_B$ configurations, respectively).}
\label{fig_experiment}
\end{figure*}

The $\lambda^\mathrm{max}$ estimated from experimental data (shown in Fig.~\ref{fig_experiment}a) are distributed around the corresponding theoretical predictions for the splay 
states.
The mean values of $\lambda^\mathrm{max}$ for the two configurations are significantly different, which we confirmed with the paired $t$-test; the null-hypothesis that the difference of the means is zero is rejected, because the $p$-value is smaller than machine precision and is certainly smaller than the significance level of $0.05$.
The difference between the two means ($\approx 0.165$) is 
substantially
larger than a typical variation of $\lambda^\mathrm{max}$ along the trajectory in a segment 
(Supplementary Fig.~\ref{fig3}a), 
indicating that our conclusion is not sensitive to uncertainty in the measurement of steady-state $\delta^*_i$ values.

In addition to establishing the statistically significant difference in the averaged $\lambda^\mathrm{max}$ between the two configurations, we investigated the
extent to which changing the configuration from $\boldsymbol{\beta}_A$ to $\boldsymbol{\beta}_B$ 
enhances the stability of the same experimentally realized states.
This was
done by recomputing $\lambda^\mathrm{max}$ for $\boldsymbol{\beta}=\boldsymbol{\beta}_A$ and $\boldsymbol{\beta}=\boldsymbol{\beta}_B$ for each time-series segment, which is justified because the steady state of Eq.~\eqref{eq-en} does not depend on the choice of $\boldsymbol{\beta}$.
Thus, the recomputed $\lambda^\mathrm{max}$ 
determines
the stability we would observe if, starting with one $\boldsymbol{\beta}$ configuration, the brakes on the generators were adjusted to realize the other configuration without changing the state.
Finding $\lambda^\mathrm{max} (\boldsymbol{\beta}_B) < \lambda^\mathrm{max} (\boldsymbol{\beta}_A) < 0$ for a given steady state implies that condition~\eqref{eqn:csb} for converse symmetry breaking is satisfied for $\boldsymbol{\beta}^* = \boldsymbol{\beta}_B$ and $\widetilde{\boldsymbol{\beta}} \approx \boldsymbol{\beta}_A$.
As shown in 
Fig.~\ref{fig_experiment}b, 
this was indeed observed to be the case for almost all the identified time-series segments (whose corresponding splay states inferred from data are shown in Fig.~\ref{fig_experiment}c).
We also verified that the observed stability improvement is robust against uncertainties in generator parameters 
(Supplementary Fig.~\ref{fig3}b).
Thus, 
our
measurements and analysis 
provide experimental evidence of converse symmetry breaking---a 
rotationally symmetric synchronous state of our system becomes stable when the rotational symmetry of the system is broken by the heterogeneity of the generator parameters.

Our demonstration of converse symmetry breaking can be interpreted from various different angles. On the one hand, it establishes a scenario in which stable symmetric states require system asymmetry. 
On the other hand, it shows that the 
assumption that increasing uniformity across individual entities would facilitate uniform behavior 
is generally 
false, even when
they are indistinguishable in terms
of their interactions.
In network systems, this equates to stating that there are situations in which the nodes need to be 
suitably
non-identical in order for them to stably converge to identical dynamical states even if 
all nodes occupy structurally equivalent positions in the network.
This leads to the counter-intuitive conclusion that node heterogeneity across a network can 
help---rather than inhibit---convergence
to a homogeneous dynamical state, as required for numerous processes including synchronization, consensus, and herding.

But what enables converse symmetry breaking?
Symmetry breaking 
itself
can be immediately appreciated by considering a one-dimensional Mexican hat potential, where the state with reflection symmetry (the system's symmetry) is unstable and the two stable states are asymmetric. This simple example illustrates the fundamental tenet that a symmetric model can describe asymmetric observations. Conversely, the phenomenon demonstrated here shows that symmetric observations may require an asymmetric model.
While potentially less immediate to visualize,
which might be the reason 
it
was not 
demonstrated
earlier,
converse symmetry breaking 
can be interpreted as arising from the following trade-off: system asymmetry 
can reduce
the likelihood of having a symmetric solution, but it can also increase the likelihood of having one such solution stable. 
This realization opens the door for new control approaches  
to manipulate system parameters and optimize the stability of symmetric states in networks whose function benefits from the symmetry of these states.
While we purposely designed our experiment with identically coupled 
oscillators
to isolate the phenomenon, the opportunities for optimization and control are even more evident when the identical coupling constraint is 
lifted, since the stabilizing effect of breaking the system symmetry with node heterogeneity is expected to be common in such cases.
We thus suggest that the results presented here will naturally extend to 
real
systems with tunable node parameters, such as
networks of logic gates, neuronal systems, coupled lasers, and networks of mechanical, electrical, and chemical  oscillators.

\bigskip\bigskip\bigskip
\noindent{\bf\large Methods}

\paragraph{Experimental design.}

The AC generators used in the experiment are of $2.5$\,V and $0.5$\,W because they are easily available and provide the desirable low-voltage, low-power output.
While larger generators could synchronize with less noise as they are machined with higher precision relative to their sizes, they require power electronics, and leave less room for errors that could cause physical damage.
To ensure that 
the
generators are 
as identical as possible, 
we obtained 
eight
generators from the same manufacturer, and selected 
three 
of them with the closest terminal voltage for a given rotational speed.
The procedure we used to measure the generator parameters is detailed in 
Supplementary Information,
Sec.~\ref{si-gendyn}.

We chose a specific AC frequency, $\omega_\mathrm{s} = 100$\,Hz, which equals $100/3 \approx 33.3$ rotations per second for the generator shaft (our generators have three pole pairs). 
The generators provide $1.55$\,V terminal (r.m.s.) voltage at this speed (without load), and we verified that the voltage was proportional to speed, up to $\approx\!55$ rotations per second. 
However, we found that the terminal voltage dropped to $\approx\!1.4$\,V when the generator was connected to the network with a typical configuration we considered.
We thus used $1.4$\,V as the base (a reference value) for normalizing voltages in all per-unit calculations.
The impedance base ($3.48\,\Omega$) was determined in the process of making a choice of electrical component parameters, as described below, and the power base ($0.56$\,W) was determined accordingly from the chosen voltage and impedance bases.

For any given combination of capacitance, inductance, and resistance that make up a rotationally symmetric circuit of the same form as in Fig.~\ref{fig_setup}b (and our choice of $\omega_\mathrm{s} = 100$\,Hz), the theoretical prediction of $\lambda^\mathrm{max}$ for the splay states for any $\boldsymbol{\beta}$ can be computed using the procedure described
in the section `Calculation of $\lambda^\mathrm{max}$' below.
The specific combination shown in Fig.~\ref{fig_setup}b was identified through multiple iterations of system design in which we made refinements in our choice of electrical components, construction of the experimental modules, and measurement apparatus.
The refinement of component parameters was performed
using their normalized, per-unit values (for example, the capacitance was chosen
such that the shunt susceptance of the $\pi$ model representing each link is $\approx\!1.04$ in per unit).
Ultimately, we chose a parameter combination in which the predicted stability of the splay states is sufficiently strong for $\widetilde{\boldsymbol{\beta}}$ and improves significantly when changing to $\boldsymbol{\beta}_g$, while ensuring that system components operate within their rated capacities and the $\boldsymbol{\beta}$ change is experimentally realizable by friction adjustment.
The actual inductors ($0.32$~H and $7.8\,\Omega$, according to the manufacturer's specification) and capacitors ($475\,\mu$F) were selected from those that were readily available to closely match the values in Fig.~\ref{fig_setup}b.

\paragraph{Modeling dynamical noise.}

We analyze the stability of Eq.~\eqref{eq-en} in the presence of the noise term $\eps\xi_i(t)$, described either by a discrete-time model or a continuous-time model.
In the discrete-time model, $\eps\xi_i(t)$ is modeled as random impulse perturbations (i.e., as a sum of Dirac delta functions with random magnitudes located at random times).
We can account for any distributions of perturbation magnitudes and times that are bounded in the sense that there is a maximum magnitude $M$ for the impulse perturbations and a minimum time interval $\tau$ between consecutive perturbations (which corresponds to a maximum rate at which the system is perturbed).
When the system is close to a splay state with maximum Lyapunov exponent $\lambda^{\max} < 0$, the dynamics approximately follows the linearized equation (as described in the main text) and is characterized by an exponential convergence to the splay state at a rate of $\lambda^{\max}$ between any two consecutive perturbations.
Thus, if the deviation of the system state from the splay state immediately after the $k$th perturbation at time $t_k$ is $\Delta\mathbf{x}(t_k)$, the deviation after the next perturbation is given by $\Delta\mathbf{x}(t_{k+1}) = e^{\lambda^{\max}\Delta t_k} \cdot \Delta\mathbf{x}(t_k) + \boldsymbol{\xi}_{k+1}$, where $\Delta t_k$ is the (random) time interval between the $k$th and $(k+1)$th perturbations, and $\boldsymbol{\xi}_k$ is the (random) displacement resulting from the $k$th impulse perturbation.
Since $\Delta t_k \ge \tau$ and $\lVert\boldsymbol{\xi}_k\lVert \le M$, we have $\lVert\Delta\mathbf{x}(t_{k+1})\rVert \le e^{\tau\lambda^{\max}} \lVert\Delta\mathbf{x}(t_k)\rVert + M$.
By recursively applying this inequality, we obtain 
\begin{equation}
\begin{split}
\lVert\Delta\mathbf{x}(t_k)\rVert &\le (e^{\tau\lambda^{\max}})^k \lVert\Delta\mathbf{x}(t_0)\rVert + M\cdot\frac{1 - (e^{\tau\lambda^{\max}})^{k+1}}{1 - e^{\tau\lambda^{\max}}}\\
&\to \frac{M}{1 - e^{\tau\lambda^{\max}}} \quad \text{as $k \to \infty$,}
\end{split}
\end{equation}
which indicates that the deviation from the splay state is bounded by $M/(1 - e^{\tau\lambda^{\max}})$ in the limit of large $k$ (and thus large $t$).
Using $\lVert\Delta\mathbf{x}(t_k)\rVert \le \Delta_\text{sync}$, $\forall k$, where $\Delta_\text{sync}$ is a constant, we see that the system would stay synchronized in the splay state if
\begin{equation}\label{eqn:criterion-noisy}
\frac{M}{1 - e^{\tau\lambda^{\max}}} < \Delta_\text{sync}.
\end{equation}
It follows that the splay state is stable in the presence of noise if the noise magnitude $M$ is sufficiently small to satisfy this condition.
On the other hand, if the asymptotic bound $M/(1 - e^{\tau\lambda^{\max}})$ is larger than $\Delta_\text{sync}$, there is a non-zero probability that the deviation $\lVert\Delta\mathbf{x}(t_k)\rVert$ exceeds $\Delta_\text{sync}$ for sufficiently large $k$, implying that the splay state is unstable.
Therefore, given $M$, $\tau$, and $\Delta_\text{sync}$, there is a (negative) threshold $\lambda^{\max}$ value
\begin{equation}
\lambda^{\max}_\text{th} \equiv \frac{1}{\tau}\ln(1 - M/\Delta_\text{sync})
\end{equation}
corresponding to the stability transition for the noisy system: the splay state is stable in the presence of noise if $\lambda^{\max} < \lambda^{\max}_\text{th}$ and unstable if $\lambda^{\max} > \lambda^{\max}_\text{th}$.
Note that each value of $\lambda^{\max}_\text{th}$ represents a range of combinations of $M$, $\tau$, and $\Delta_\text{sync}$.
In general, for any pair of configurations $\boldsymbol{\beta}_\text{I}$ and $\boldsymbol{\beta}_\text{II}$ with $\lambda^{\max}(\boldsymbol{\beta}_\text{I}) < \lambda^{\max}(\boldsymbol{\beta}_\text{II}) < 0$, there is a range of combinations of $M$, $\tau$, and $\Delta_\text{sync}$ for which $\lambda^{\max}(\boldsymbol{\beta}_\text{I}) < \lambda^{\max}_\text{th} < \lambda^{\max}(\boldsymbol{\beta}_\text{II})$, i.e., the splay state is stable for the $\boldsymbol{\beta}_\text{I}$ configuration, while it is unstable for the $\boldsymbol{\beta}_\text{II}$ configuration.

In the continuous-time model, the term $\eps\xi_i(t)$ represents the effect of Brownian noise on the dynamics of the $i$th oscillator.
Since the effect of the noise decays most slowly along the eigenmode associated with the eigenvalue $\lambda^{\max}<0$, we focus on the projection of the full six-dimensional dynamics onto that eigenmode.
We model this projected dynamics by the following stochastic differential equation:
\begin{equation}\label{eqn:lambda-max-model}
dx(t) = \lambda^{\max} \, x(t)dt + \sigma d\xi(t),
\end{equation}
where $x(t)$ represents the (one-dimensional) deviation from the frequency-synchronous state, $\xi(t)$ is the standard Brownian noise (i.e., $\xi(t) - \xi(0)$ for fixed $t$ follows the Gaussian distribution with zero mean and variance $t$), and $\sigma>0$ is a constant that can be used to tune the level of noise intensity felt by the system.
The stochastic process given by Eq.~\eqref{eqn:lambda-max-model} is called the Ornstein--Uhlenbeck (OU) process~\cite{Uhlenbeck:1930,Gillespie:1991}.
Given the constants $\sigma$ and $\Delta_\text{sync}$, we define $\text{Pr}(\lambda^{\max})$ to be the probability that $|x(t)| \le \Delta_\text{sync}$ for $t \to \infty$ (i.e., the probability that the system is synchronized in the large $t$ limit).
From the known property of the OU process~\cite{Doob:1942} that $x(t)$ for a fixed $t$ follows the Gaussian distribution with mean $e^{t\lambda^{\max}}x(0)$ and variance $\sigma^2(1-e^{2t\lambda^{\max}})/(-2\lambda^{\max})$, we compute $\text{Pr}(\lambda^{\max}) = \text{erf}(\Delta_\text{sync}\sqrt{-\lambda^{\max}}/\sigma)$, where $\text{erf}$ denotes the error function.
Note that, as $\lambda^{\max} < 0$ is increased (and thus $\sqrt{-\lambda^{\max}}$ is decreased), the probability $\text{Pr}(\lambda^{\max})$ decreases.
Thus, for a given constant $p_\text{th}$ (chosen close to one), we define the stability threshold $\lambda^{\max}_\text{th}$ as the value of $\lambda^{\max}$ for which $\text{Pr}(\lambda^{\max}) = p_\text{th}$.
It then follows that $\lambda^{\max}_\text{th} = -\sigma^2[\text{erf}^{-1}(p_\text{th})]^2/\Delta_\text{sync}^2$.
Note that the threshold $\lambda^{\max}_\text{th} = \lambda^{\max}_\text{th}(\sigma)$ depends on the noise intensity level $\sigma$.
For $\sigma=0$ (noiseless case), we have $\lambda^{\max}_\text{th}(0) = 0$, recovering the stability threshold for deterministic systems.
As $\sigma$ increases, the threshold $\lambda^{\max}_\text{th}(\sigma)$ monotonically decreases, indicating that the decay of deviation needs to be faster to maintain the same probability $\text{Pr}(\lambda^{\max})$ for higher noise levels.
Therefore, if $\lambda^{\max}(\boldsymbol{\beta}_\text{I}) < \lambda^{\max}(\boldsymbol{\beta}_\text{II}) < 0$, there is a range of $\sigma$ for which $\lambda^{\max}(\boldsymbol{\beta}_\text{I}) < \lambda^{\max}_\text{th}(\sigma) < \lambda^{\max}(\boldsymbol{\beta}_\text{II})$, i.e., the system stays near the frequency-synchronous state with high probability as $t\to\infty$ for the $\boldsymbol{\beta}_\text{I}$ configuration, while it does not for the $\boldsymbol{\beta}_\text{II}$ configuration.

\paragraph{Matrix $\mathbf{P}$ for splay states.}

Assuming ideal system components, the electric circuit we use (Fig.~\ref{fig_setup}b) is rotationally symmetric, and thus the complex 
node-to-node
admittances $Y_{0ik}$ are identical for all $i \neq k$.
If, in addition, the internal impedances $z_{\mathrm{int},i}$ of the three generators are identical, then the effective admittances $Y_{ik}$ are also identical, i.e., $Y_{ik} = y = |y| \exp(j\alpha)$ for all $i \neq k$, where $j$ is the imaginary unit.
Assuming further that the inertia constants $H_i$ are all
identical, the steady splay states satisfy $c_{ik} = c$ for all $i \neq k$, and the matrix $\mathbf{P}$ in Eq.~\eqref{eqn:P} takes the form
\begin{equation}
\begin{pmatrix}
-b-b' & b & b' \\
b' & -b-b' & b \\
b & b' & -b-b'
\end{pmatrix},
\end{equation}
where 
$b = c \bigl( -\frac{1}{2}\cos\gamma \mp \frac{\sqrt{3}}{2} \sin\gamma \bigr)$, $b' = c \bigl(-\frac{1}{2}\cos\gamma \pm \frac{\sqrt{3}}{2} \sin\gamma \bigr)$, and $\gamma = \alpha - \pi/2$.
The topology of the underlying network corresponds to one favoring converse symmetry breaking in the analysis of coupled oscillators presented in Ref.~\citen{Yuanzhao17}.

\paragraph{Calculation of $\boldsymbol{\lambda^\mathrm{max}}$.}

For both the theoretically predicted stability landscape in Fig.~\ref{fig_setup} and the estimates from the experimental measurements in Fig.~\ref{fig_experiment} and 
Supplementary Fig.~\ref{fig3}, 
we compute $\lambda^\mathrm{max}$ using the following procedure.
We first apply the Kirchoff law to compute the active and reactive power injections at the terminal of each generator ($P_i$ and $Q_i$, respectively) from the network's inductances and capacitances (at a given frequency) as well as the terminal voltage magnitudes and angles ($|V_i|$ and $\theta_i$, respectively).
We then apply the Kirchoff law again to the internal impedance of each generator (assuming the classical model) to compute $|E_i|$ and $\delta_i^*$ from $P_i$, $Q_i$, $|V_i|$, and $\theta_i$.
From these and the values of $H_i$, the parameters $c_{ik}$ and $\gamma_{ik}$ can be computed.
We can then obtain $\lambda^\mathrm{max}$ for any given $\boldsymbol{\beta}$ by computing the eigenvalues of $\mathbf{J} = \bigl( \begin{smallmatrix}\,\,\,\,\mathbf{0} & \,\,\,\,\mathbf{1}\\ -\mathbf{P} & -\mathbf{B}\end{smallmatrix}\bigr)$ with the matrix $\mathbf{P}$ defined in Eq.~\eqref{eqn:P}.
For the theoretical prediction, we use $\omega_\mathrm{s} = 100$\,Hz and the terminal voltage magnitude $|V_i|=1.4$\,V.
We set the angles $\theta_i$ to have exactly $120^{\circ}$ differences, which ensures splay-state values for $\delta_i$ as we assume the generator parameters to be identical: $r_{\mathrm{int},i} = 1.6 \,\Omega$ and $x_{\mathrm{int},i} = 3.34 \,\Omega$ for all $i$.
We compute $H_i$ from $J_i = 5 \times 10^{-5}$ kg$\cdot$m$^2$, which makes $H_i$ all identical.
For the experimental estimation of $\lambda^\mathrm{max}$ from a given time-series segment, we use the average of the measured values over the segment for $\omega_\mathrm{s}$, $|E_i|$, and $\delta_i^*$, along with the values of $J_i$, $r_{\mathrm{int},i}$, and $x_{\mathrm{int},i}$ measured for each generator (see 
Supplementary Information, Secs.~\ref{supp-inertia-meas} and \ref{supp-meas-impedance}
for our measurement procedure).

\paragraph{Conducting the experiment and taking measurements.}

Arduino Uno and Arduino Mega microcontrollers were used to provide sufficient computational capacity and I/O channels to facilitate measurement and control of the experiment.
The Arduino Uno has six analog inputs, 
of which three were used to measure the terminal voltages of the three generators, and the remaining three were used to measure the currents passing through the generator terminals.
The same controller also has several digital outputs, four of which were used to control relays that switch the voltage inputs between measurement signals and test signals.
The test signals were used to calibrate the voltage levels.
We used three analog inputs on the Arduino Mega to take shaft rotational speed measurement from infrared sensors that detect markings on the shafts. 
This capability was used in measuring $\beta_i$ for each generator (see 
Supplementary Information, Sec.~\ref{supp-beta-meas}
for details).
In addition, three digital outputs of this controller were used to drive relays that switch the generators on and off in the AC circuit.

Each experimental run was performed as follows.
First, while the generator terminal relays were 
open (i.e., disconnected from the circuit),
we turned on the DC motors driving the generators, and we adjusted their speed  
manually to 
$100/3 \approx 33.3$
rotations per second, 
yielding 
terminal voltage signals with frequencies approximately at the desired $100$\,Hz (noting that the frequency is three times higher than the shaft's rotational frequency because our generators have three pole pairs).
Once 
this
was achieved, the generator terminal relays were closed, establishing the coupling between the generators.
This can result in a transient voltage dynamics with large fluctuations, with the speed of each generator dropping due to the impedance of the connected network.
The speed of each DC motor was adjusted once again to ensure that each generator was running at the desired speed. When the last generator approached the desired speed, the synchronous state formed spontaneously. Real-time voltage phasor readouts were displayed on the computer interface to confirm that the system had achieved a desired splay state. At this point, we started recording the voltage phasor readings.

External perturbations from vibrations and other sources continuously disturb
the synchro\-nous state, and the resulting transients prevented us from observing this state for more than a few seconds at a time. Moreover, as a result of components heating up during the experiment, the amount of friction between the shaft and its housing tends to change at different rates for
different generators, leading to unbalanced net changes in the mechanical power input of the generators that distort the splay state. We mitigated this problem by re-adjusting the speed of the DC motors. These adjustments are implemented manually, aided by the real-time display of the phase angle differences of the generators.  To prevent damage due to rising temperature of the components, we were limited to a maximum of 10 min of continuous run when no brakes were applied (the $\boldsymbol{\beta}_A$ configuration), and a maximum of 5 min when brakes were applied (the $\boldsymbol{\beta}_B$ configuration).

The terminal voltage was recorded at $3{,}320$ samples per second for each generator using the analog-to-digital converters of the Arduino Uno, along with the microsecond-accurate time stamps from the microcontroller's internal timer.
The data were streamed to the computer interface for post-processing in which the raw readings were converted to time-dependent 
frequency and phasor angle and magnitude 
with the original resolution. 
This conversion was done using custom software for phasor measurement, which implements the following calculations. 
For each 
data point, 
we took
the last three oscillatory periods of 
the voltage signal
and calculated the least-squares fit of a sinusoidal function with its amplitude, offset, phase shift, and frequency as the fitting parameters.  
The resulting amplitude, phase shift, and frequency were used as estimates of the terminal voltage peak magnitude, phasor angle, and frequency, respectively, for that data point.
To eliminate measurement noise from the terminal voltage phasor and frequency time series, we applied a $3^\text{rd}$-order Savitzky-Golay filter~\cite{Orfanidis:1996}, with a window size of $0.4$\,s.
The generators' internal voltage phasors $E_i = |E_i| \exp(j \delta_i)$ were then obtained using the Kirchoff laws, along with the calculated 
terminal voltage phasors, the instantaneous synchronous frequency (computed as the average of the instantaneous AC frequencies of the generator terminals),
and each component's capacitance and inductance.

\paragraph{Identifying time-series segments of splay states.}

Because some deviation from the 
exact 
splay state is unavoidable, the steady-state phase angles $\delta_i^*$ are not necessarily separated by exactly $120$ degrees.
We 
thus
identified in each time series the set of maximal segments that satisfy the following criteria: 1) for each data point in the segment, one generator is ahead by an angle $\Delta^{+}$ and another
is behind by $\Delta^{-}$ relative to the other generator (taken to be generator 1 here), while satisfying $|\Delta^{\pm} - 120^\circ| < 10^\circ$; and 2) the length of the segment is at least $0.1$\,s (approximately ten cycles of voltage).
In total, we have obtained
275
segments for the $\boldsymbol{\beta}_A$ configuration and 
190
segments for the $\boldsymbol{\beta}_B$ configuration.
Supplementary Fig.~\ref{fig:window-lengths} 
shows the length distribution of these segments, which are within the timescale for which Eq.~\eqref{eq-en} is valid.
Within these segments, we verified that the generator frequencies were synchronized: the maximum (instantaneous) frequency difference among the three generators
was $<1$\,Hz, and the standard deviation of each generator's frequency was $<0.25$\,Hz in $93$\%
of the
$465=275+190$
identified segments.
We also verified 
that all parameters in the deterministic part of Eq.~\eqref{eq-en} were constant within experimental noise (see 
Supplementary Information, Sec.~\ref{app:eq-en-derivation} and Supplementary Fig.~\ref{fig:validation} 
for details).
This further confirmed the validity of Eq.~\eqref{eq-en} for describing the $\delta_i$ dynamics in each time-series segment.

\paragraph{Data Availability.}
All data that support results in this article are available from the corresponding author upon reasonable request.

\paragraph{Code Availability.}
The custom code used for the analysis of the data from the experiment is available from the corresponding author upon reasonable request.


\clearpage

\bigskip\noindent{\bf\large Acknowledgements}

\noindent
The authors thank John B.~Ketterson for insightful discussions about this research.
This research was funded by ARO Grant No. W911NF-15-1-0272 and
by Northwestern University's Finite Earth Initiative (supported by Leslie and Mac
McQuown).

\bigskip\noindent{\bf\large Author contributions}

\noindent
F.M., T.N., and A.E.M. designed the research and contributed to the modeling.  F.M. performed the experiments and simulations.  F.M., T.N., and A.E.M. analyzed the results and wrote the paper.  All authors approved the final manuscript.

\bigskip\noindent{\bf\large Competing interests}

\noindent
The authors declare no competing interests.

\bigskip\noindent{\bf\large Materials \& Correspondence}

\noindent
Correspondence and material requests should be addressed to T.N.\ or A.E.M.\\
(E-mails: t-nishikawa@northwestern.edu or motter@northwestern.edu).

\clearpage


\baselineskip18pt

\setcounter{equation}{0}
\renewcommand{\theequation}{S\arabic{equation}}

\bigskip\noindent{\bf\Large Supplementary Information}

\noindent {\it\small Network experiment demonstrates converse symmetry breaking}

\renewcommand{\thesection}{S\arabic{section}}
\renewcommand{\thesubsection}{S\arabic{section}.\arabic{subsection}}

\setcounter{section}{0}

\section{Derivation of coupled oscillator network model}
\label{app:eq-en-derivation}

Our derivation of the deterministic part of the model in Eq.~\eqref{eq-en} of the main text is based on the so-called classical model of a 
generator.
For completeness, we first reproduce a derivation of the classical model
starting 
from Newton's second law written in terms of the total torque accelerating the rotor of generator~$i$:
\begin{equation}
J_i \ddot{\phi}_{\mathrm{m},i} = - d_i \dot{\phi}_{\mathrm{m},i} + T_{\mathrm{m},i} - T_{\mathrm{e},i},
\label{eq-newton1}
\end{equation}
where $\phi_{\mathrm{m},i}$ is the mechanical (rotor shaft) angle of generator $i$ (relative to a stationary axis), $J_i$ is the total moment of inertia of the rotor, $T_{\mathrm{m},i}$ is the mechanical torque provided to the rotor, $T_{\mathrm{e},i}$ is the electrical torque (load), and $d_i$ is the damping-torque coefficient accounting for windage and friction.
Changing to a frame of reference rotating at the synchronous (mechanical) frequency $\omega_\mathrm{sm}$ of the rotor through $\widetilde{\phi}_{\mathrm{m},i} \equiv \phi_{\mathrm{m},i} - \omega_\mathrm{sm}  t$, Eq.~\eqref{eq-newton1} becomes
\begin{equation}
J_i \ddot{\widetilde{\phi}}_{\mathrm{m},i} = - d_i (\dot{\widetilde{\phi}}_{\mathrm{m},i} + \omega_\mathrm{sm}) + T_{\mathrm{m},i} - T_{\mathrm{e},i}.
\label{eq-newton2}
\end{equation}
Multiplying this by the angular velocity $\omega_{\mathrm{m},i} \equiv \dot{\phi}_{\mathrm{m},i}$, we can write
\begin{equation}
\begin{split}
\omega_{\mathrm{m},i} J_i \ddot{\widetilde{\phi}}_{\mathrm{m},i} &= - \omega_{\mathrm{m},i} d_i (\dot{\widetilde{\phi}}_{\mathrm{m},i} + \omega_\mathrm{sm}) + \omega_{\mathrm{m},i} T_{\mathrm{m},i} - \omega_{\mathrm{m},i} T_{\mathrm{e},i}\\
 &= - \omega_{\mathrm{m},i} d_i (\dot{\widetilde{\phi}}_{\mathrm{m},i} + \omega_\mathrm{sm}) + P_{\mathrm{m},i} - P_{\mathrm{e},i},
\end{split}
\end{equation}
where $P_{\mathrm{m},i}$ is the mechanical power supplied to the rotor and $P_{\mathrm{e},i}$ is the electrical power drawn from the rotor. 
When the system is close to a frequency-synchronous state, we have $\omega_{\mathrm{m},i} \approx \omega_\mathrm{sm}$, and we can write
\begin{equation}
\omega_\mathrm{sm} J_i \ddot{\widetilde{\phi}}_{\mathrm{m},i} = -\omega_\mathrm{sm} d_i \dot{\widetilde{\phi}}_{\mathrm{m},i} + P_{\mathrm{m},i} - \omega_\mathrm{sm}^2 d_i - P_{\mathrm{e},i}.
\label{eq-swing1}
\end{equation}
Thus, in a synchronous steady state in which $\dot{\widetilde{\phi}}_{\mathrm{m},i} = \ddot{\widetilde{\phi}}_{\mathrm{m},i} = 0$, the mechanical power input must balance the electrical power 
output 
plus all the losses due to damping and friction. Dividing Eq.~\eqref{eq-swing1} by a power base ($P_{\mathrm{base}}$) allows us to express power in per-unit quantities:
\begin{equation}
\frac{\omega_\mathrm{sm} J_i}{P_{\mathrm{base}}} \, \ddot{\widetilde{\phi}}_{\mathrm{m},i} = -\frac{\omega_\mathrm{sm} d_i}{P_{\mathrm{base}}} \, \dot{\widetilde{\phi}}_{\mathrm{m},i} + P_{\mathrm{m},i}^{\mathrm{(pu)}} - \frac{\omega_\mathrm{sm}^2 d_i}{P_{\mathrm{base}}} - P_{\mathrm{e},i}^{\mathrm{(pu)}},
\end{equation}
which leads to 
\begin{equation}
\frac{2H_i}{\omega_\mathrm{sm}} \, \ddot{\widetilde{\phi}}_{\mathrm{m},i} = - \frac{D_i}{\omega_\mathrm{sm}} \, \dot{\widetilde{\phi}}_{\mathrm{m},i} + \widetilde{P}_{\mathrm{m},i}^{\mathrm{(pu)}} - P_{\mathrm{e},i}^{\mathrm{(pu)}},
\end{equation}
where $H_i \equiv  \frac{1}{2} J_i \omega_\mathrm{sm}^2 / P_{\mathrm{base}}$ is the inertia constant (which equals the kinetic energy of the rotor at the synchronous speed), $D_i \equiv d_i \omega_\mathrm{sm}^2 / P_{\mathrm{base}}$ is the damping coefficient, and  $\widetilde{P}_{\mathrm{m},i}^{\mathrm{(pu)}} \equiv P_{\mathrm{m},i}^{\mathrm{(pu)}} - \omega_\mathrm{sm}^2 d_i / P_{\mathrm{base}}$ is the net power input.

We assume that
each generator $i$ can be
represented by two nodes: an internal node, with voltage phasor $E_i = |E_i| \exp(j \delta_i)$, whose magnitude $|E_i|$ is assumed to be constant; and the terminal node, whose voltage is $V_i = |V_i| \exp(j \theta_i)$ and matches the generator's terminal voltage. Note that $j$ denotes the imaginary unit.
The internal and terminal nodes are connected through an (internal) impedance, which we measured in our experiment (see Sec.~\ref{supp-meas-impedance} below).
If the generator has $p$ pole pairs, the mechanical and electrical angles are related by $\delta_i = p \widetilde{\phi}_{\mathrm{m},i}$ (and thus $\dot{\delta}_i = p  \dot{\widetilde{\phi}}_{\mathrm{m},i}$, $\ddot{\delta}_i = p  \ddot{\widetilde{\phi}}_{\mathrm{m},i}$, and the synchronous voltage frequency $\omega_\mathrm{s} = p \omega_\mathrm{sm}$), leading to a second form of the
equation of motion:
\begin{equation}
\frac{2H_i}{\omega_\mathrm{s}} \ddot{\delta}_i = - \frac{D_i}{\omega_\mathrm{s}} \dot{\delta}_i + \widetilde{P}_{\mathrm{m},i}^{\mathrm{(pu)}} - P_{\mathrm{e},i}^{\mathrm{(pu)}}.
\label{eq-swing2}
\end{equation}
This is equivalent to the classical model.
The voltage at the terminal node
is related to the terminal voltages of the other generators by the Kirchoff laws as 
\begin{equation}\label{eqn:rpf}
P_i = \sum_{k=1}^{n} |V_i V_k Y_{0ik}| \sin (\theta_i - \theta_k - \alpha_{0ik} + \pi/2),
\end{equation} 
where $P_i$ is the real power injection from generator $i$ into the network, 
and $Y_{0ik} = |Y_{0ik}| \exp(j \alpha_{0ik})$ are the complex entries of 
the admittance matrix $\mathbf{Y}_0 = (Y_{0ik})$ representing the AC electric network.
Through a procedure known as Kron reduction applied to the network (see, for example, Refs.~\citen{And:03} and \citen{Nish:15} of the main text), we can also express
the real power injection $P_{\mathrm{e},i}^{\mathrm{(pu)}}$ at 
the internal node of generator $i$
in a form similar to Eq.~\eqref{eqn:rpf}, but in terms of the internal voltage angles $\delta_i$:
\begin{equation}
P_{\mathrm{e},i}^{\mathrm{(pu)}} = \sum_{k=1}^n |E_i E_k Y_{ik}| \sin (\delta_i - \delta_k - \alpha_{ik} + \pi/2),
\end{equation}
where $\mathbf{Y} = (Y_{ik})$ is the {\it effective} admittance matrix with $Y_{ik} = |Y_{ik}| \exp(j \alpha_{ik})$, representing the effective coupling between generators. 
Substituting this into Eq.~\eqref{eq-swing2} and defining
\begin{equation}\label{eqn:parameters}
\begin{split}
\beta_i &\equiv \frac{D_i}{2H_i} = \frac{d_i}{J_i},\\
a_i &\equiv \frac{\omega_\mathrm{s} \bigl[\widetilde{P}_{\mathrm{m},i}^{\mathrm{(pu)}} - |E_i^2 Y_{ii}| \cos (\alpha_{ii}) \bigr]}{2H_i},\\
c_{ik} &\equiv \frac{\omega_\mathrm{s}|E_i E_k Y_{ik}|}{2H_i},\\
\gamma_{ik} &\equiv \alpha_{ik} - \pi/2,
\end{split}
\end{equation}
we obtain the 
oscillator network model in the form of Eq.~\eqref{eq-en} of the main text (with $\eps=0$).

Note that $\beta_i$ is constant because $d_i$ and $J_i$ are both physical constants characterizing the corresponding generator.
Note also that, for a synchronous state with a given (constant) frequency $\omega_\mathrm{s}$, the parameter $H_i = \frac{1}{2} J_i \omega_\mathrm{sm}^2 / P_{\mathrm{base}} = \frac{1}{2} J_i \omega_\mathrm{s}^2 / (p^2 P_{\mathrm{base}})$ and the admittances $Y_{ik} = |Y_{ik}| \exp(j \gamma_{ik})$ are constant.
Thus, if the parameters $\widetilde{P}_{\mathrm{m},i}^{\mathrm{(pu)}}$ and $E_i$ are constant, then all the parameters defined in Eq.~\eqref{eqn:parameters} would be constant.
In our experiment, we validated the constancy of $\omega_\mathrm{s}$, $\widetilde{P}_{\mathrm{m},i}^{\mathrm{(pu)}}$, and $E_i$ directly from the measurements for each time-series segment 
(Supplementary Fig.~\ref{fig:validation}).
The linearization of Eq.~\eqref{eq-en} involving the matrix $\mathbf{P}$ in Eq.~\eqref{eqn:P} assumes that all parameters in Eq.~\eqref{eqn:parameters} are constant, which is valid when $\omega_\mathrm{s}$, $\widetilde{P}_{\mathrm{m},i}^{\mathrm{(pu)}}$, and $E_i$ are constant.

\section{Measurement of generator parameters}
\label{si-gendyn}

We experimentally measured all generator parameters required for our analysis, i.e., the parameter $\beta_i$, the moment of inertia 
$J_i$, 
and the internal impedance $z_{\mathrm{int},i}$ for each generator ($i=1,2,3$). 
To simplify the notations, the generator index $i$ in the subscript are omitted in the following subsections.

\subsection{Effective damping parameter $\boldsymbol{\beta}$}
\label{supp-beta-meas}

The adjustable parameter $\beta$ for each generator combines the moment of inertia and all damping 
effects
for the given generator, 
$\beta = d/J$. 
When both the mechanical and the electrical torques are turned off, Eq.~\eqref{eq-newton1} becomes 
$\ddot{\phi}_{\mathrm{m}} = -\frac{d}{J} \dot{\phi}_{\mathrm{m}} = -\beta \dot{\phi}_{\mathrm{m}}$,
which can be expressed as $\dot{\omega}_{\mathrm{m}} = -\beta \omega_{\mathrm{m}}$.
Thus, the rotor decelerates at an exponential rate of 
$\beta$. 
We can directly measure this rate by fitting an exponential decay curve to the experimentally measured rotor speed, after the generator is disconnected from the circuit and 
the DC motor driving it
is turned off.
We thus equipped the rotating shaft of each generator with a reflective infrared phototransistor that detects markings on the shaft, providing direct measurements of the rotor frequency.
The function we used
to fit the frequency measurement $\omega_\mathrm{m}(t)$ 
is the following:
\begin{align}
y(t) = \begin{cases}
        A, & t < t_0,\\
        A \exp (-\beta t), &  t_0 \leq t < t_1,
        \end{cases}
\end{align}
where the steady-state speed $A$, the turn-off time $t_0$, and the rate of deceleration $\beta$ are fitting parameters.
The turn-off time $t_0$ needs to be fitted because the switching is initiated manually rather than by the microcontroller.
The measured values of $\omega_\mathrm{m}(t)$ were taken up to $t=t_1$, at which $\omega_\mathrm{m}(t) \approx 20$ Hz; 
below this threshold the exponential decay is not a good approximation.

A typical fitting scenario
is shown in Supplementary Fig.~\ref{si-fig-betas}a, and 
a single set of
measurements of $\beta$ in the $\boldsymbol{\beta}_A$ and $\boldsymbol{\beta}_B$ configurations are shown in 
panels b--d of the same figure.
The standard deviation of these measurements was $\approx 0.1$, indicating the precision of the $\beta$ values that we could configure and confirm by 
experimental measurements.
At the beginning of each experimental run, we 
adjusted
the breaks if necessary to ensure that the average of 
$\beta$ (estimated from at least ten measurements)
was within $\pm 0.1$ of the corresponding designed value for the given $\boldsymbol{\beta}$ configuration.

\subsection{Moment of inertia $\boldsymbol{J}$}
\label{supp-inertia-meas}

The measurement of $J$ is based on the measurement of $\beta$, which was performed by estimating the exponential rate of deceleration as the DC motor driving the generator was turned off, while no load was attached to the generator.
If we repeat this experiment but with a load connected to the terminal of the generator, then the generator would decelerate slightly faster.
Using a resistor with $R = 2\,\Omega$ as the load (rated for a maximum of $10$~W dissipation)
would result in a time-dependent electric torque, 
\begin{equation}
T_\mathrm{e}(t) =  \frac{P_\mathrm{e}(t)}{\omega_\mathrm{m}(t)} = \frac{V^2(t)}{R \omega_\mathrm{m}(t)},
\end{equation}
where $V(t)$ is the instantaneous r.m.s.\ voltage measured across the generator terminals.
Substituting into Eq.~\eqref{eq-newton1} and using 
$\omega_\mathrm{m} = \dot{\phi}_\mathrm{m}$ and $T_\mathrm{m} = 0$,
we obtain
\begin{equation}
\dot{\omega}_\mathrm{m} = -\beta \omega_\mathrm{m} - \frac{V^2(t)}{JR \omega_\mathrm{m}(t)},
\end{equation}
whose solution is given by
\begin{equation}
\omega_\mathrm{m} (t) = \exp \left( -\beta t \right) \left( \omega_\mathrm{m}^2(0) - \frac{2}{J} \int_{0}^{t}\frac{V^2(t')}{R} \exp(2 \beta t') dt' \right)^{\frac{1}{2}}.
\label{eq-inertia-integral}
\end{equation}
With this and the measured time series of $V(t)$, we can calculate $\omega_\mathrm{m}(t)$ numerically for any $t$.
The parameter $J$ can then be estimated by minimizing the sum of squared differences between these calculated values of $\omega_\mathrm{m}(t)$ and the corresponding direct measurements for all $t$ (taken up to a time at which $\omega_\mathrm{m}(t) \approx 20$\,Hz).

The necessary voltage 
values $V(t)$
for Eq.~\eqref{eq-inertia-integral} were provided by direct measurement of the terminal voltages, which were taken simultaneously with the rotor speed 
measurements (using a reflective infrared phototransistor, as was done for the $\beta$ measurements; see Sec.~\ref{supp-beta-meas} above).
Since
the voltage measurements were taken at a much higher rate than the rotor speed measurements ($3{,}320$ samples per second for the voltage vs.\ $480$ samples per second for the rotor speed at $\omega_\mathrm{m} = 40\times 2\pi$~rad/s, 
which decreases as the rotor slows down), 
we used the sampling rate of the voltages for the time discretization of the integral in Eq.~\eqref{eq-inertia-integral}.
For the $\beta$ value in Eq.~\eqref{eq-inertia-integral}, we substituted the estimate from Sec.~\ref{supp-beta-meas} above, which is an average over many measurements (see, for example, Supplementary Fig.~\ref{si-fig-betas}).  The estimate of $J$ for each generator was then obtained by repeating this procedure several times (using the same average $\beta$ value) and taking the average.  A typical fitting curve, as well as the estimated values of $J$ are provided in Supplementary Fig.~\ref{si-fig-inertia}.

\subsection{Internal impedance $\boldsymbol{z_\mathrm{int}}$}
\label{supp-meas-impedance}

The measurement of the generator's internal impedance $z_\mathrm{int} = r_\mathrm{int} + jx_\mathrm{int}$ requires 
the open-circuit voltage ($V_\mathrm{oc}$) and the short-circuit current ($I_\mathrm{sc}$) to be measured at the same rotor speed. 
We can take these readings directly using the voltage and current sensors.
Our experimental setup has a set of ACS712 current sensors that can measure the current passing through each generator's terminal.
In
large industrial synchronous generators the rotating magnetic field is 
induced by an adjustable field current, thus 
making the $V_\mathrm{oc}$ and $I_\mathrm{sc}$ values dependent on this current as well. 
Our permanent-magnet generators, however, have constant magnetic field, and thus $V_\mathrm{oc}$ and $I_\mathrm{sc}$ are functions of the rotor speed only.
Both $V_\mathrm{oc}$ and $I_\mathrm{sc}$ generally increase linearly with the speed until the air gap flux starts to saturate the iron parts of the generator.
This linear regime was observed up to $55$ rotations per second for our generators.
We performed all our experiments and measurements in this regime.

The resistive component $r_\mathrm{int}$ of the internal impedance in the classical model is negligible for a typical synchronous generator, but we found that it was non-negligible for our generators.
We measured both the resistive and reactive components using the following procedure.
We first obtained the magnitude of the internal impedance using the relation $|z_\mathrm{int}| = \frac{V_\mathrm{oc}}{I_\mathrm{sc}}$, with $V_\mathrm{oc}$ and $I_\mathrm{sc}$ measured at $40$ rotations per second.
We then measured the real part $r_\mathrm{int}$ directly by connecting a multimeter to the generator terminals in resistance measurement mode.
The remaining imaginary part of the internal impedance was calculated using $x_\mathrm{int} = \sqrt{|z_\mathrm{int}|^2 - r_\mathrm{int}^2}$. 
The measured values of $V_\mathrm{oc}$, $I_\mathrm{sc}$, and $r_\mathrm{int}$, as well as the calculated $x_\mathrm{int}$, are as follows:
\begin{itemize}
    \item {\bf Generator 1}: $V_\mathrm{oc}=1.85$\,V,  $I_\mathrm{sc}=0.50$\,A, $r_\mathrm{int}=1.6$\,$\Omega$, $x_\mathrm{int}=3.34$\,$\Omega$.
    \item {\bf Generator 2}: $V_\mathrm{oc}=1.90$\,V,  $I_\mathrm{sc}=0.51$\,A, $r_\mathrm{int}=1.6$\,$\Omega$, $x_\mathrm{int}=3.36$\,$\Omega$.
    \item {\bf Generator 3}: $V_\mathrm{oc}=1.79$\,V,  $I_\mathrm{sc}=0.50$\,A, $r_\mathrm{int}=1.6$\,$\Omega$, $x_\mathrm{int}=3.20$\,$\Omega$.
\end{itemize}
The mean and the standard deviation among the three generators are $V_\mathrm{oc}=1.85 \pm 0.055$\,V,  $I_\mathrm{sc}=0.50 \pm 0.058$\,A, $r_\mathrm{int}=1.6 \pm 0.0$\,$\Omega$, and $x_\mathrm{int}=3.30 \pm 0.087$\,$\Omega$.
Note that the measured internal impedances of our generators would be called the synchronous impedance in the power systems literature, since $I_\mathrm{sc}$ is measured at a steady speed.
This 
quantity is usually used to model the steady-state synchronous dynamics after a short period of transients following a short circuit, 
while the so-called transient impedance is used to model the transients.
For our generators, however, we found no detectable transient, implying that the transient impedance is essentially equal to the synchronous impedance.
Therefore, we used the $z_\mathrm{int}$ measured by the above procedure in our analysis of short-term dynamics.

\clearpage


\hoffset 0cm
\textwidth 16cm
\bigskip\noindent{\bf\Large Supplementary Figures}

\newcounter{sfigure}
\renewcommand{\figurename}{Supplementary Fig.}
\renewcommand{\thefigure}{\arabic{sfigure}}

\addtocounter{sfigure}{1} 
\vspace{5mm}
\begin{figure}[ht]
\center
\includegraphics[width=\columnwidth]{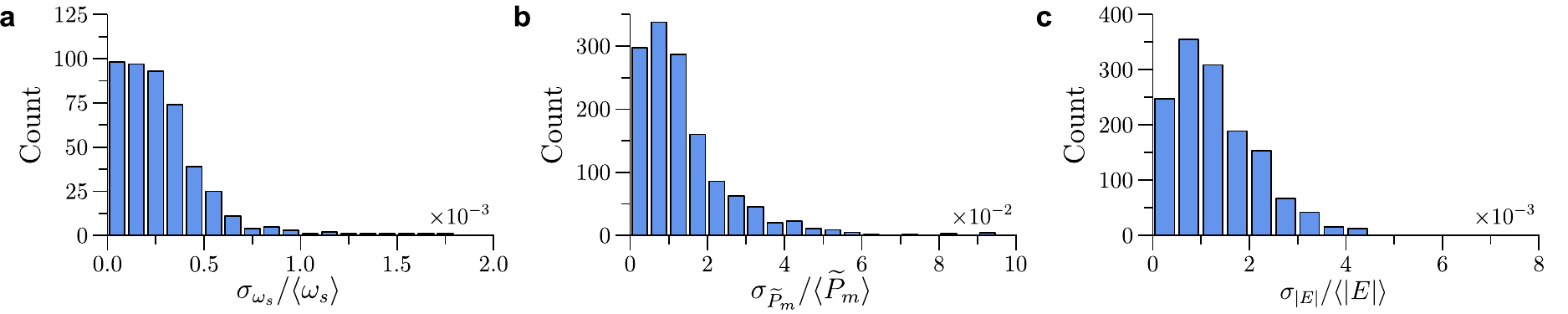}
\caption{\small
{\bf Validation of model assumptions.}
{\bf a}--{\bf c}, Histograms of normalized fluctuations for the parameters $\omega_\mathrm{s}$ ({\bf a}), $\widetilde{P}_{\mathrm{m},i}^{\mathrm{(pu)}}$ ({\bf b}), and $E_i$ ({\bf c}) 
(defined in 
Sec.~\ref{app:eq-en-derivation}).
We quantify experimental fluctuations of a given parameter $q$ of a given generator in a given time-series segment by $\sigma_q/\langle q \rangle$, where $\sigma_q$ and $\langle q \rangle$ are the standard deviation and average, respectively, of the measured instantaneous values of $q$ across the segment.
The histograms, taken over all generators and over all time-series segments for each parameter, 
indicate that the (relative) magnitude of fluctuations in these parameters is approximately constant in each time-series segment, validating the corresponding assumptions underlying Eq.~\eqref{eq-en} of the main text.}
\label{fig:validation}
\end{figure}

\addtocounter{sfigure}{1} 
\vspace{5mm}
\begin{figure}[ht]
\center
\includegraphics[scale=1.2]{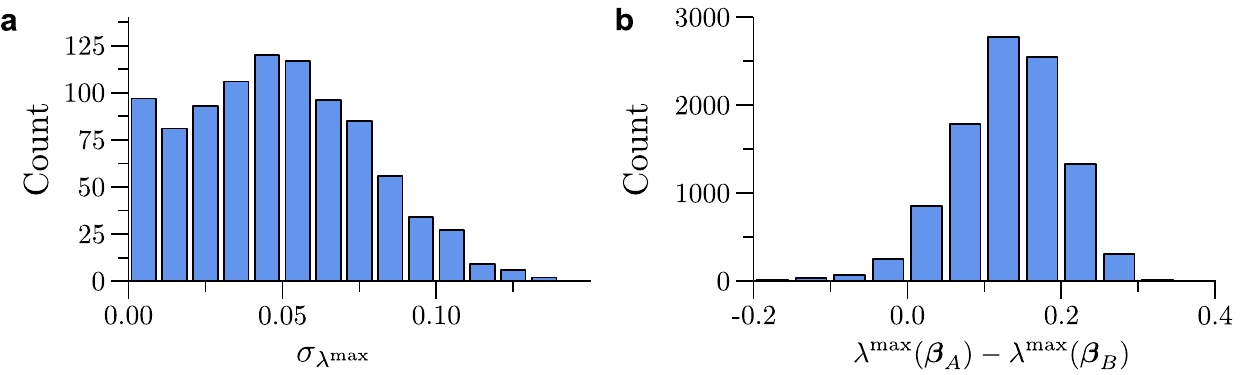}
\vspace{-2mm}
\caption{\small 
{\bf Robustness of measured stability improvement.}
{\bf a}, Distribution of $\sigma_{\lambda^{\mathrm{max}}}$, the standard deviation of $\lambda^{\mathrm{max}}$ as it varies along the trajectory in a given segment, where $\lambda^{\mathrm{max}}$ is computed using the instantaneous $\delta^*_i$ values at each data point in the segment.
{\bf b}, Histogram of predicted stability improvement 
$\lambda^{\mathrm{max}}(\boldsymbol{\beta}_{A}) - \lambda^{\mathrm{max}}(\boldsymbol{\beta}_{B})$ 
when the dynamical parameters of the generators are randomly perturbed (using $10{,}000$ realizations).
The specific parameters perturbed were $V_\mathrm{oc}$, $I_\mathrm{sc}$, and $J$ for each generator 
(defined in 
Secs.~\ref{app:eq-en-derivation}, \ref{supp-inertia-meas}, and \ref{supp-meas-impedance}).
The perturbations were drawn from the normal distribution having the mean and the standard deviation of the measured values of a given parameter of each generator (see 
Sec.~\ref{supp-meas-impedance} and Supplementary Fig.~\ref{si-fig-inertia}).
The vast majority of the realizations lead to a measurable stability difference between the two 
$\boldsymbol{\beta}$ 
configurations, showing that 
converse symmetry breaking
is observable and robust even under realistic 
uncertainties in the state and parameters of the system.}
\label{fig3}
\end{figure}

\addtocounter{sfigure}{1}
\begin{figure}[ht]
\center
\includegraphics[width=0.51\columnwidth]{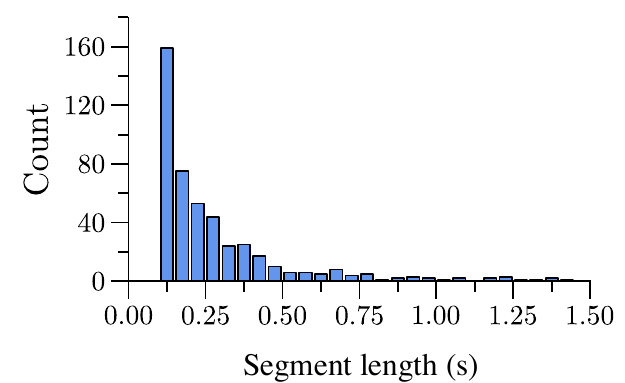}
\caption{\small
{\bf Length distribution of time-series segments.}
The histogram shows the distribution of the lengths of the time-series segments used in the analysis shown in Fig.~\ref{fig_experiment} (for both the 
$\boldsymbol{\beta}_{A}$ and $\boldsymbol{\beta}_{B}$ 
configurations).}
\label{fig:window-lengths}
\end{figure}

\addtocounter{sfigure}{1}
\begin{figure}[ht]
\center
\includegraphics[scale=1]{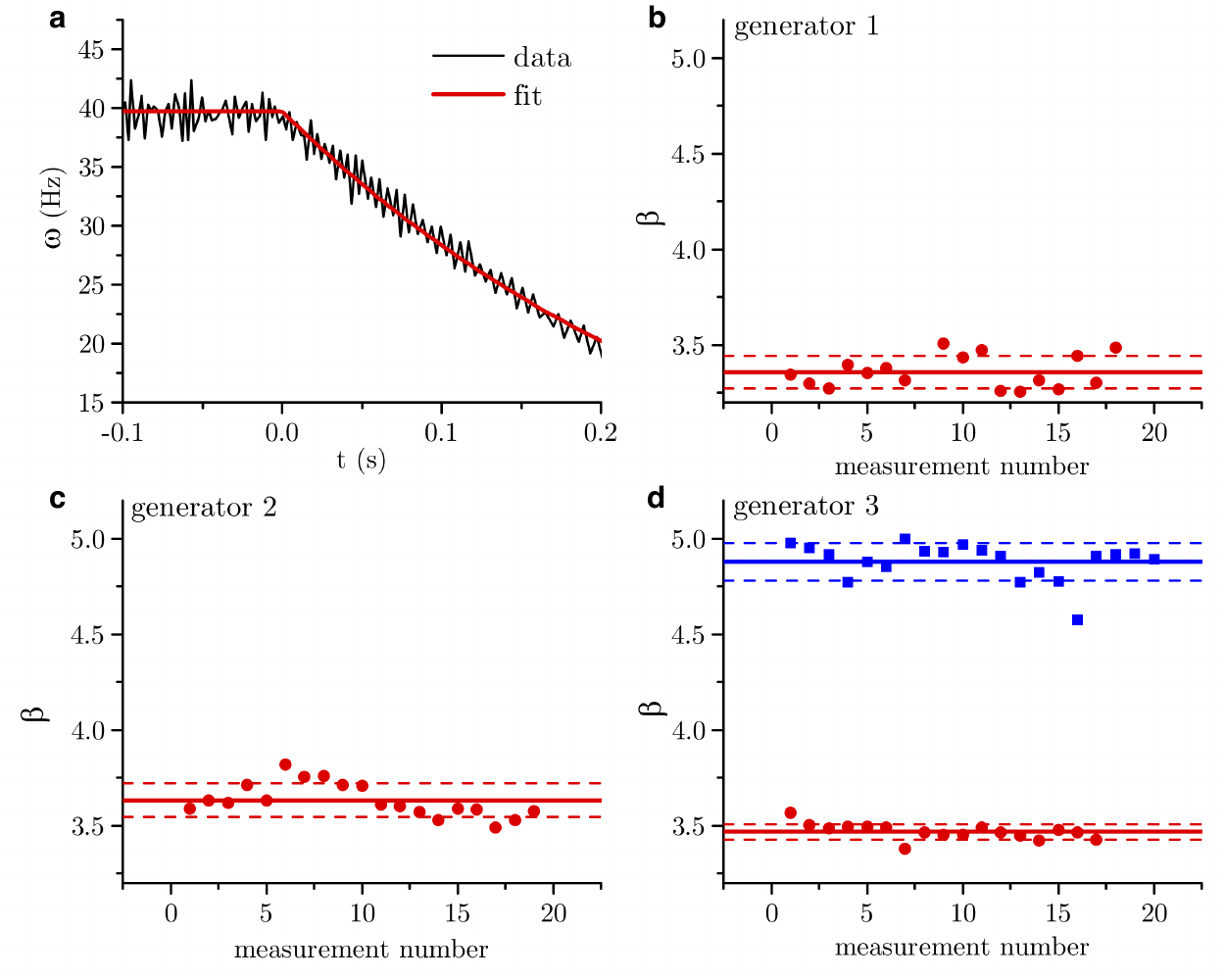}
\caption{\small
{\bf Measurement of parameter $\boldsymbol{\beta}$.} {\bf a}, Typical fitting scenario, where 
the steady-state frequency, the turn-off time (shifted to $t=0$), and the rate of deceleration $\beta$ 
are fitted to the data from rotor frequency measurement.
{\bf b},{\bf c}, Measurements of the $\beta$ value for generators 1 and 2, respectively. {\bf d}, Measurements of the $\beta$ values for generator 3 in the uniform 
($\boldsymbol{\beta}_A$, 
red dots) and non-uniform 
($\boldsymbol{\beta}_B$, 
blue squares) configurations. In {\bf b}--{\bf d}, the mean and standard deviation of the data points are indicated by the solid and dashed lines, respectively.}
\label{si-fig-betas}
\end{figure}

\addtocounter{sfigure}{1}
\begin{figure}[ht]
\center
\includegraphics[scale=1]{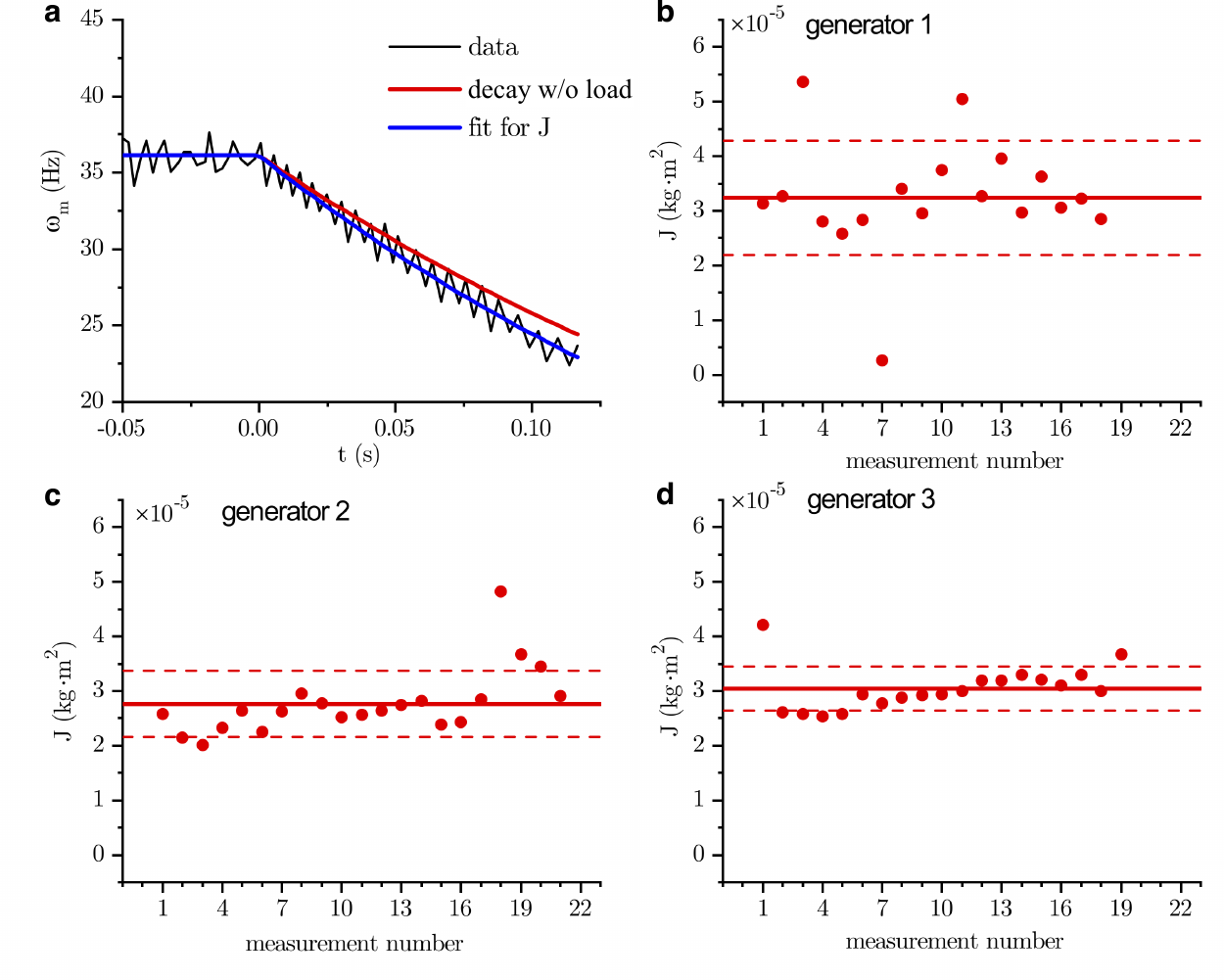}
\caption{\small
{\bf Measurement of parameter $\boldsymbol{J}$.} {\bf a}, 
Decay of the measured rotor frequency from a typical run with a resistive load of $2\,\Omega$, along with the corresponding fit of the steady-state frequency, turn-off time (shown as $t=0$ here), and parameter $J$. 
We also show as a reference the predicted decay without the load.
{\bf b}--{\bf d},~Measurements of $J$ for generators 1, 2, and 3, respectively. In {\bf b}--{\bf d}, the mean and standard deviation of the data points are indicated by the solid and dashed lines, respectively.}
\label{si-fig-inertia}
\end{figure}


\begin{thebibliography}{10}

\bibitem{Pikovsky:01}
Pikovsky, A., Rosenblum, M. \& Kurths, J.
{\it Synchronization: A Universal Concept in Nonlinear Sciences}
(Cambridge University Press, 
Cambridge,  
2001).

\bibitem{rev5}
Arenas, A., D\'{i}az-Guilera, A., Kurths, J., Moreno, Y. \& Zhou, C.
Synchronization in complex networks. 
{\it Phys. Rep.} {\bf 469}, 93--153 (2008).

\bibitem{rev3}
D\"orfler, F. \& Bullo, F.
Synchronization in complex networks of phase oscillators: A survey.
{\it Automatica} {\bf 50}, 1539--1564 (2014).

\bibitem{Pecora:15}
Pecora, L. M. \& Carroll, T. L.
Synchronization of chaotic systems.
{\it Chaos} {\bf 25}, 097611 (2015).

\bibitem{Yamaguchi:03}
Yamaguchi, S. et al.
Synchronization of cellular clocks in the suprachiasmatic nucleus.
{\it Science} {\bf 302}, 1408--1412 (2003).

\bibitem{nat1}
Yamaguchi, Y. et al.
Mice genetically deficient in vasopressin V1a and V1b Receptors are resistant to jet lag.
{\it Science} {\bf 342}, 85--90 (2013).

\bibitem{Lu:16}
Lu, Z. et al.
Resynchronization of circadian oscillators and the east-west asymmetry of jet-lag.
{\it Chaos} {\bf 26}, 094811 (2016).

\bibitem{Ranta:95}
Ranta, E., Kaitala, V., Lindstr\"{o}m, J. \& Linden, H.
Synchrony in population dynamics.
{\it Proc. R. Soc. Lond. B} {\bf 262}, 113--118 (1995).

\bibitem{Schwartz:02}
Schwartz, M. K., Mills, L. S., McKelvey, K. S., Ruggiero, L. F. \& Allendorf, F. W.
DNA reveals high dispersal synchronizing the population dynamics of Canada lynx.
{\it Nature} {\bf 415}, 520--522 (2002).

\bibitem{McClintock:71}
McClintock, M. K.
Menstrual synchrony and suppression.
{\it Nature} {\bf 229}, 244--245 (1971).

\bibitem{Strogatz:05}
Strogatz, S. H., Abrams, D. M., McRobie, A., Eckhardt, B. \& Ott, E.
Theoretical mechanics: crowd synchrony on the Millennium Bridge.
{\it Nature} {\bf 438}, 43--44 (2005).

\bibitem{eng1}
Rosin, D. P., Rontani, D., Gauthier, D. J. \& Sch\"oll, E. 
Control of synchronization patterns in neural-like Boolean networks.
{\it Phys. Rev. Lett.} {\bf 110}, 104102 (2013).

\bibitem{Fischer:06}
Fischer, I. et al.
Zero-lag long-range synchronization via dynamical relaying.
{\it Phys. Rev. Lett.} {\bf 97}, 123902 (2006).

\bibitem{eng2}
Zamora-Munt, J., Masoller, C., Garc\'{i}a-Ojalvo, J. \& Roy, R.
Crowd synchrony and quorum sensing in delay-coupled lasers.
{\it Phys. Rev. Lett.} {\bf 105}, 264101 (2010).

\bibitem{Argyris:16}
Argyris, A., Bourmpos, M. \& Syvridis, D.
Experimental synchrony of semiconductor lasers in coupled networks.
{\it Opt. express} {\bf 24}, 5600--5614 (2016).

\bibitem{Kiss:02}
Kiss, I. Z., Zhai, Y. \& Hudson, J. L.
Emerging coherence in a population of chemical oscillators.
{\it Science} {\bf 296}, 1676--1678 (2002).

\bibitem{eng3}
Kiss, I. Z., Rusin, C. G., Kori, H. \& Hudson, J. L.
Engineering complex dynamical structures: sequential patterns and desynchronization.
{\it Science} {\bf 316}, 1886--1889 (2007).

\bibitem{Fon:17}
Fon, W. et al.
Complex dynamical networks constructed with fully controllable nonlinear nanomechanical oscillators.
{\it Nano Lett.} {\bf 17}, 5977--5983 (2017).

\bibitem{Hill:06}
Hill, D. J. \& Chen, G.
Power systems as dynamic networks.
In {\it Proceedings of the 2006 IEEE International Symposium on Circuits and Systems},
722--725 
(2006).

\bibitem{Mot:13}
Motter, A. E., Myers, S. A., Anghel, M. \& Nishikawa, T.
Spontaneous synchrony in power-grid networks.
{\it Nat. Phys.} {\bf 9}, 191--197 (2013).

\bibitem{Dorfler:13}
D\"{o}rfler, F., Chertkov, M. \& Bullo, F.
Synchronization in complex oscillator networks and smart grids.
{\it Proc. Natl. Acad. Sci. USA} {\bf 110}, 2005--2010 (2013).

\bibitem{TakashiAIS}
Nishikawa, T. \& Motter, A. E.
Symmetric states requiring system asymmetry.
{\it Phys. Rev. Lett.} {\bf 117}, 114101 (2016).

\bibitem{Okuda:91}
Okuda, K. \& Kuramoto, Y.
Mutual entrainment between populations of coupled oscillators. 
{\it Prog. Theor. Phys.} {\bf 86}, 1159--1176 (1991).

\bibitem{Golubitsky:1988}
Golubitsky, M., Stewart, I., \& Schaeffer, D. G.
{\it Singularities and groups in bifurcation theory}, Vol. 2
(Springer-Verlag, New York, 1988).

\bibitem{Golubitsky:1999}
Golubitsky, M., \& Stewart, I.
Symmetry and pattern formation in coupled cell networks.
In {\it Pattern Formation in Continuous and Coupled Systems},
65--82 (Springer-Verlag, New York, 1999).

\bibitem{Nicosia:13}
Nicosia, V., Valencia, M., Chavez, M., D\'{i}az-Guilera, A. \& Latora, V.
Remote synchronization reveals network symmetries and functional modules.
{\it Phys. Rev. Lett.} {\bf 110}, 174102 (2013).

\bibitem{Pecora:14}
Pecora, L. M., Sorrentino, F., Hagerstrom, A. M., Murphy, T. E. \& Roy, R. 
Cluster synchronization and isolated desynchronization in complex networks with symmetries.
{\it Nat. Commun.} \textbf{5}, 4079 (2014).

\bibitem{Whalen:15}
Whalen, A. J., Brennan, S. N., Sauer, T. D. \& Schiff, S. J.
Observability and controllability of nonlinear networks: the role of symmetry.
{\it Phys. Rev. X} {\bf 5}, 011005 (2015).

\bibitem{Sorrentino:16}
Sorrentino, F., Pecora, L. M., Hagerstrom, A. M., Murphy, T. E. \& Roy, R.
Complete characterization of the stability of cluster synchronization in complex dynamical networks.
{\it Sci. Adv.} {\bf 2}, e1501737 (2016).

\bibitem{Zhang:17}
Zhang, L., Motter, A. E. \& Nishikawa, T.
Incoherence-mediated remote synchronization.
{\it Phys. Rev. Lett.} {\bf 118}, 174102 (2017).

\bibitem{Cho:17}
Cho, Y. S., Nishikawa, T. \& Motter, A. E.
Stable chimeras and independently synchronizable clusters. 
{\it Phys. Rev. Lett.} {\bf 119}, 084101 (2017).

\bibitem{Barrett:17}
Barrett, W., Francis, A. \& Webb, B.
Equitable decompositions of graphs with symmetries.
{\it Linear Algebra Appl.} {\bf 513}, 409--434 (2017).

\bibitem{MacArthur:08}
MacArthur, B. D., S\'{a}nchez-Garc\'{i}a, R. J. \& Anderson, J. W.
Symmetry in complex networks.
{\it Discrete Appl. Math.} {\bf 156}, 3525--3531 (2008).

\bibitem{ch1}
Kuramoto, Y. \& Battogtokh, D. 
Coexistence of coherence and incoherence in nonlocally coupled phase oscillators.
{\it Nonlinear Phenom. Complex Syst.} {\bf 5}, 380--385 (2002).

\bibitem{ch2}
Abrams, D. M. \& Strogatz, S. H.
Chimera states for coupled oscillators.
{\it Phys. Rev. Lett.} {\bf 93}, 174102 (2004).

\bibitem{Chimera3}
Panaggio, M. J. \& Abrams, D. M.
Chimera states: coexistence of coherence and incoherence in networks of coupled oscillators.
{\it Nonlinearity} {\bf 28}, R67 (2015).

\bibitem{ch3}
Hagerstrom, A. M. et al.
Experimental observation of chimeras in coupled-map lattices.
{\it Nat. Phys.} {\bf 8}, 658--661 (2012).

\bibitem{ch4}
Tinsley, M. R., Nkomo, S. \& Showalter, K.
Chimera and phase-cluster states in populations of coupled chemical oscillators.
{\it Nat. Phys.} {\bf 8}, 662--665 (2012).

\bibitem{Martens:13}
Martens, E. A., Thutupalli, S., Fourri\`ere, A. \& Hallatschek, O.
Chimera states in mechanical oscillator networks.
{\it Proc. Natl. Acad. Sci. USA} {\bf 110}, 10563--10567 (2013).

\bibitem{ChimeraXP3}
Hart, J. D., Bansal, K., Murphy, T. E. \& Roy, R.
Experimental observation of chimera and cluster states in a minimal globally coupled network.
{\it Chaos} {\bf 26}, 094801 (2016).

\bibitem{Yuanzhao17}
Zhang, Y., Nishikawa, T. \& Motter, A. E.
Asymmetry-induced synchronization in oscillator networks.
{\it Phys. Rev. E} {\bf 95}, 062215 (2017).

\bibitem{Yuanzhao2018}
Zhang, Y. \& Motter, A. E.
Identical synchronization of nonidentical oscillators: when only birds of different feathers flock together.
{\it Nonlinearity} {\bf 31}, R1 (2017).

\bibitem{Grainger:1994}
Grainger, J. \& Stevenson, W.
{\it Power System Analysis}.
(McGraw-Hill Co., Singapore, 1994).

\bibitem{And:03}
Anderson, P. M. \& Fouad, A. A. 
{\it Power System Control and Stability}
(IEEE Press, 
Piscataway, 
2003).

\bibitem{Nish:15}
Nishikawa, T. \& Motter, A. E.
Comparative analysis of existing models for power-grid synchronization.
{\it New J. Phys.} {\bf 17}, 015012 (2015).

\bibitem{pg3}
Susuki, Y., Mezi\'c, I. \& Hikihara, T. 
Coherent swing instability of power grids.
{\it J. Nonlinear Sci.} {\bf 21}, 403--439 (2011).

\bibitem{pg2}
Lozano, S., Buzna, L. \& D\'{i}az-Guilera, A.
Role of network topology in the synchronization of power systems,
{\it Eur. Phys. J. B} {\bf 85}, 231--238 (2012).

\bibitem{Men:14}
Menck, P. J., Heitzig, J., Kurths, J. \& Schellnhuber, H. J.
How dead ends undermine power grid stability.
{\it Nat. Commun.} {\bf 5}, 3969 (2014).

\bibitem{Auer:16}
Auer, S., Kleis, K., Schultz, P., Kurths, J. \& Hellmann, F.
The impact of model detail on power grid resilience measures.
{\it Eur. Phys. J. Special Topics} {\bf 225}, 609--625 (2016).

\bibitem{Schafer:18}
Sch\"{a}fer, B., Beck, C., Aihara, K., Witthaut, D. \& Timme, M.
Non-Gaussian power grid frequency fluctuations characterized by L\'{e}vy-stable laws and superstatistics.
{\it Nat. Energy} {\bf 3}, 119--126 (2018).

\bibitem{r1} 
Burke, J. V., Lewis, A. S. \& Overton, M. L. 
A robust gradient sampling algorithm for nonsmooth, nonconvex optimization. 
{\it SIAM J. Optimiz.} {\bf 15}, 751--779 (2005).

\bibitem{r2} 
Freitas, P. \& Lancaster, P. 
On the optimal value of the spectral abscissa for a system of linear oscillators. 
{\it SIAM J. Matrix Anal. A.} {\bf 21}, 195--208 (1999).

\bibitem{r3} 
Kirillov, O. N. \& Overton, M. L. 
Robust stability at the swallowtail singularity. 
{\it Frontiers in Physics} {\bf 1}, 1--9 (2013).

\bibitem{r4}
Boyd, S. 
Convex optimization of graph Laplacian eigenvalues. 
In {\it Proc. ICM} {\bf 3}, 1311--1319 (2006).

\bibitem{r5}
De Abreu, N. M. M. 
Old and new results on algebraic connectivity of graphs. 
{\it Linear Algebra Appl.} {\bf 423}, 53--73 (2007).

\bibitem{splay}
Zou, W. \& Zhan, M.
Splay states in a ring of coupled oscillators: from local to global coupling.
{\it SIAM J. Appl. Dyn. Syst.} {\bf 8}, 1324--1340 (2009).

\bibitem{Uhlenbeck:1930}
Uhlenbeck, G. E. \& Ornstein, L. S.
On the theory of the Brownian motion.
{\it Phys. Rev.} {\bf 36}, 823--841 (1930).

\bibitem{Gillespie:1991}
Gillespie, D. T.
{\it Markov Processes: An Introduction for Physical Scientists}
(Academic Press, Inc., San Diego, 1991).

\bibitem{Doob:1942}
Doob, J. L. 
The Brownian movement and stochastic equations.
{\it Ann. Math.}, Vol. 43, 351--369 (1942).

\bibitem{Orfanidis:1996}
Orfanidis, S. J.
{\it Introduction to Signal Processing}
(Prentice Hall, Englewood Cliffs,
1996).


\end{thebibliography}
\end{document}